\documentclass{article}
\usepackage{authblk}
\usepackage{graphicx}%
\usepackage{multirow}%
\usepackage{amsmath,amssymb,amsfonts}%
\usepackage{amsthm}%
\usepackage{mathrsfs}%
\usepackage[title]{appendix}%
\usepackage[english,greek]{babel}
\usepackage{xcolor}%
\usepackage{textcomp}%
\usepackage{manyfoot}%
\usepackage{booktabs}%
\usepackage{algorithm}%
\usepackage{algorithmicx}%
\usepackage{algpseudocode}%
\usepackage{listings}%
\usepackage{placeins}
\usepackage{makecell}
\usepackage{hyperref}
\usepackage[super,sort&compress]{natbib}

\begin{document}
\selectlanguage{english}

\begin{titlepage}
\centering

\title{Enhancing Hole Mobility in Monolayer {WSe\textsubscript{2}}\ p-FETs via Process-Induced Compression}

\author[1,4,5,6]{{He Lin} {Zhao}}

\author[1,5,6]{{Sheikh Mohd Ta-Seen} {Afrid}}

\author[3,4,6]{{Dongyoung} {Yoon}}

\author[2,4,6]{{Zachary} {Martin}}

\author[3,4,6]{{Zakaria} {Islam}}

\author[5,6,7]{{Sihan} {Chen}}

\author[3,4,6]{{Yue} {Zhang}}

\author[2,4,6]{{Pinshane Y.} {Huang}}

\author[1,5,6]{{Shaloo} {Rakheja}}

\author[1,2,3,4,5,6]{{Arend M.} {van der Zande} \thanks{Corresponding author contact: {arendv@illinois.edu}}}

\affil[1]{{Department of Electrical and Computer Engineering}, {University of Illinois Urbana-Champaign}, {{Urbana}, {61801}, {Illinois}, {United States}}}

\affil[2]{{Department of Materials Science and Engineering}, {University of Illinois Urbana-Champaign}, {{Urbana}, {61801}, {Illinois}, {United States}}}

\affil[3]{{Department of Mechanical Science and Engineering}, {University of Illinois Urbana-Champaign}, {{Urbana}, {61801}, {Illinois}, {United States}}}

\affil[4]{{Materials Research Laboratory}, {University of Illinois Urbana-Champaign}, {{Urbana}, {61801}, {Illinois}, {United States}}}

\affil[5]{{Holonyak Micro and Nanotechnology Laboratory}, {University of Illinois Urbana-Champaign}, {{Urbana}, {61801}, {Illinois}, {United States}}}

\affil[6]{{Grainger College of Engineering}, {University of Illinois Urbana-Champaign}, {{Urbana}, {61801}, {Illinois}, {United States}}}

\affil[7]{{Microelectronics Thrust}, {The Hong Kong University of Science and Technology (Guangzhou)}, Guangzhou, 511453, Guangdong, China}

\renewcommand\Affilfont{\small}
\date{}
\maketitle

\end{titlepage}

\begin{abstract}
Understanding the interactions between strain, interfacial mechanics, and electrical performance is critical for designing beyond silicon electronics based on hetero-integrated 2D materials. Through combined experiment and simulation, we demonstrated and analyzed the enhancement of hole mobility in p-type monolayer {WSe\textsubscript{2}} field effect transistors (FETs) under biaxial compression. 
We tracked FET performance versus strain by incrementing compressive strain to {WSe\textsubscript{2}} channels via sequential {AlO\textsubscript{x}} deposition and performing intermediate photoluminescence and transport measurements. 
The hole mobility factor increased at a rate of 340~\textpm~95 {\%/\%\textgreek{e}}, and the on-current factor increased at a rate of 460~\textpm~340 {\%/\%\textgreek{e}}. 
Simulation revealed that the enhancement under compression arises primarily from a reduction in inter-valley scattering between the \textgreek{G}--K valence bands, and the rate is robust against variations in carrier density, impurity density, or dielectric environment. 
These findings show that compressive strain is a powerful technique for enhancing performance in 2D p-FETs and that it is multiplicative with defect and doping engineering. 
\end{abstract}


\normalsize

\section*{Main}
\label{sec:introduction}
\FloatBarrier

Strain engineering has been a key strategy to increasing drain current density in silicon transistors across many technology nodes, with in-plane biaxial tensile strain enhancing n-FETs and uniaxial compressive strain along the direction of current enhancing p-FETs. \cite{strained_si_1, strained_si_2, strained_si_3} This strategy has new significance with
two-dimensional (2D) materials, especially semiconducting transition metal dichalcogenides (TMD). Due to their superior electrical performance\cite{2d_scaling_projection} and mechanical strength at the atomic limit,\cite{tsmc_first_2dgaa} monolayer 2D TMDs are now on industry roadmaps as channel materials for future beyond-silicon nanosheet transistor architectures. \cite{imec_roadmap}
However, 2D nanosheet channels often undergo significant out-of-plane sagging and corrugation as a result of residual strains developed during processing. \cite{tsmc_2d_cfet, tsmc_stacked_2dgaa} 
In parallel, the performance reported for 2D FETs exhibits wide variability. \cite{2d_fet_performance_review, 2d_fet_mobility_var} \par

Particularly, the deterministic design and fabrication of 2D p-FETs have lagged behind 2D n-FETs, \cite{pfet_challenges, wang_wse2_contact} demanding a more comprehensive understanding of the nanoscale mechanics governing how deposited thin films interact with 2D materials and their influence on electrical performance.\cite{al2o3_ndope, vacancy_doping}
The existence of recent studies that show tensile strain enhances electron mobility ($\mu_\mathrm{e}$) in monolayer {MoS\textsubscript{2}} and {WS\textsubscript{2}} n-FETs indicates that strain is an important factor in determining 2D transistor performance. 
For example, the reported mobility factor strain-tuning rates {$\left[\frac{\partial(\mu/\mu_0)}{\partial \varepsilon}\right]$} range from 86 percent per percent of strain (86{\%/\%\textgreek{e}}) to 200{\%/\%\textgreek{e}}, \cite{mos2_strain_review, shin_mos2_enh2, yang_mos2_enh}, and the highest strain-induced on-current enhancement in {MoS\textsubscript{2}} FETs reaches +100\%. \cite{yang_mos2_enh, pop_mos2_enh} 
Yet, strain is generally uncontrolled or unknown in most p-type device designs. A critical next step is to develop an equivalent understanding by isolating and unraveling the relative impact of strain, doping, and defects on 2D p-type transistor performance.

\begin{figure}
    \centering
    \includegraphics[width=1\linewidth]{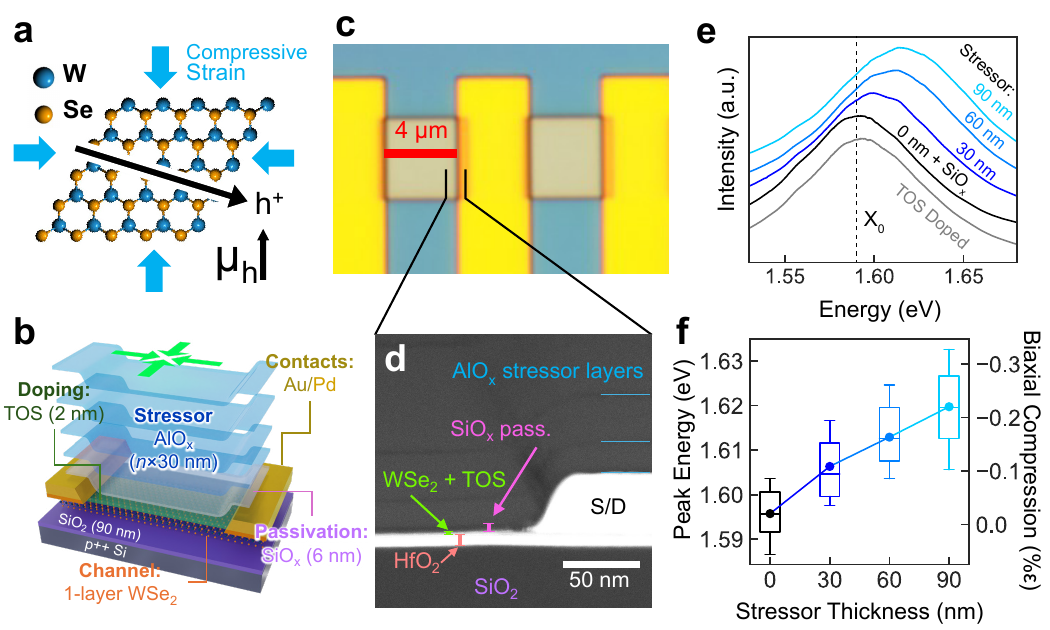}
    \caption{\textbf{Process induced strain engineering in monolayer {WSe\textsubscript{2}} p-FETs.} 
    \textbf{a,} Concept illustration of enhanced hole mobility in monolayer {WSe\textsubscript{2}} induced by biaxial compressive strain.  
    \textbf{b,} Exploded render of the important layers and experimental design in the process-strained {WSe\textsubscript{2}} FET.  \textbf{c,} Optical image of the p-FET array. \textbf{d,} a cross-sectional STEM at the contact at the position marked in (c). Each layer is annotated. \textbf{e, } Representative Photoluminescence (PL) spectra versus energy from a single {WSe\textsubscript{2}}\ channel. Grey and black spectra come from the channel after TOS doping, and 6 nm SiO\textsubscript{x} passivation. Shades of blue represent the spectra after incremental stressor depositions. \textbf{f,} PL peak energy (left) and estimated biaxial compression \textgreek{e}(right), as a function of {AlO\textsubscript{x}} thickness. Error bars represent spatially aggregated quartile statistics across 5 FET channels.}
    \label{fig:illusts_pl}
\end{figure}

Fig.~\ref{fig:illusts_pl}a illustrates the principle of our study, which is to combine experimentation and simulation to systematically investigate the role of compression on the hole mobility enhancement of monolayer {WSe\textsubscript{2}} p-FETs under process-induced compressive strain. 
We chose to study compression in anticipation that it would enhance the hole mobility of {WSe\textsubscript{2}}, mirroring the trend demonstrated in Si. As we will show, this hypothesis was correct, but the tuning rises from very different mechanisms of intervalley scattering. We report an impressive hole mobility factor strain-tuning rate under compression of {$\left[\frac{\partial(\mu/\mu_0)}{\partial \varepsilon}\right]$} of 340~\textpm~95~{\%/\%\textgreek{e}}, more than 4\texttimes\ greater than the tuning rates predicted in silicon p-FETs predicted by recent simulations.\cite{roisin_strained_si_enh} We found through simulations that the mobility enhancement factors are robust against variations in other key variables, such as carrier and impurity densities. This shows that strain-induced mobility tuning is a powerful performance multiplier in 2D CMOS.
\par

\FloatBarrier
\section*{Experimental design of compressed monolayer\\ {WSe\textsubscript{2}} p-FETs}
\FloatBarrier

Fig.~\ref{fig:illusts_pl}b illustrates how we systematically induce compressive strain in monolayer {WSe\textsubscript{2}} p-FETs. 
We adapted a method we previously developed to study the influence of process-induced tensile-strained {MoS\textsubscript{2}} FETs,  \cite{yue_mos2_enh}, using the deposition of high-stress thin films to induce strain in the underlying 2D monolayer p-FET. \cite{pena_mos2_strain, azizi_mos2_strain, azizi_graphene_strain} 
We incrementally deposit {AlO\textsubscript{x}} via e-beam evaporation upon arrays of monolayer {WSe\textsubscript{2}} FETs. E-beam evaporation deposited {AlO\textsubscript{x}}\ is a transparent amorphous film with a residual tensile stress of 0.4 GPa \textpm\ 10 MPa at room temperature; consequently, upon relaxation, the {AlO\textsubscript{x}} applies a compressive strain to the underlying 2D monolayer. See Methods and Extended Data Fig.~\ref{fig:ex_wafer_bend} for {AlO\textsubscript{x}} stress analysis. We henceforth refer to the {AlO\textsubscript{x}} film as the ``stressor''. At each stressor thickness, we measure both the photoluminescence spectra to non-intrusively quantify the strain in the channel and the FET transport characteristics. 
This method enables us to systematically correlate the changes in the performance and transport characteristics of each transistor with the induced compression. \par

Fig.~\ref{fig:illusts_pl}c shows the optical image of the finished strained {WSe\textsubscript{2}}\ p-FETs. To enable statistical analysis, we fabricated arrays of identically sized monolayer {WSe\textsubscript{2}} p-FETs on a substrate consisting of atomic layer deposition (ALD) grown 10 nm {HfO\textsubscript{2}} on 90 nm dry SiO\textsubscript{2} on degenerately doped p-Si.
We then opened PMMA windows to locally deposit the stressors on top of the channels. 
Fig.~\ref{fig:illusts_pl}d shows a cross-sectional scanning-transmission electron microscope (STEM) image of the channel-contact junction as indicated in Fig.~\ref{fig:illusts_pl}d, where each of the layers is visible as changes in contrast.
The striated patterns in the deposited {AlO\textsubscript{x}}\ stressor layers directly visualize the incremental deposition process. \cite{yue_stressor} See Extended Data Fig.~\ref{fig:ex_fab} and Methods for the fabrication process details. \par

Here, we outline the purpose of each layer in the strained FET design. The Si-SiO\textsubscript{2} substrate serves as a global back gate that is independent of the applied strain on top of the channel. 
The bottom {HfO\textsubscript{2}} layer serves to increase the dielectric breakdown voltage and also acts as an etch-stop against XeF\textsubscript{2} gas. 
Since one major challenge in making {WSe\textsubscript{2}} p-FETs is achieving and maintaining low contact resistance throughout multiple processes, we included a tungsten oxy-selenide (TOS) p-doping layer that both maintains p-type channel behavior and forms ohmic tunneling contacts between {WSe\textsubscript{2}} and Pd.\cite{sihan_TOS, oberoi_TOS, borah_TOS_doping}
After depositing contacts and defining the PMMA stressor window, we first performed a low-power e-beam evaporation of 6 nm of {SiO\textsubscript{x}}, which serves as a passivation layer that protects the channel by maintaining the integrity of the TOS layer and the {WSe\textsubscript{2}} monolayer underneath. 
Extended Data Fig.~\ref{fig:ex_doping_pasiv} and Supplementary Discussion 1 articulate the role and choice of SiO\textsubscript{x} as the passivation layer.\par

The FET channels are 4 \textgreek{m}m long \texttimes\ 5 \textgreek{m}m wide, while the stressor window is 5 \textgreek{m}m long \texttimes\ 4.6 \textgreek{m}m wide. 
The channels need to be large relative to the spot size of the laser used for optical characterization to allow for spatially resolved strain measurements. 
The mismatched stressor and channel dimensions create a clamping geometry that facilitates efficient strain transfer into the channel and promotes uniform strain distribution. \cite{yue_mos2_enh}
Thus, we design the stressor to overlap slightly with the contact electrodes but to be recessed from the free edges of the 2D channel along the width to avoid anchoring to the substrate.

\section*{Optical {WSe\textsubscript{2}}\ strain characterization}
\label{sec:pl}

We employ optical photoluminescence (PL) spectral analysis to non-destructively extract strain ($\varepsilon$) from the channel region of the {WSe\textsubscript{2}}\ FETs. In {WSe\textsubscript{2}}\, the PL neutral exciton ($X_0$) peak energy shifts with strain at a rate that depends on the strain type: 120 meV/\% for biaxial strain and 54 meV/\% for uniaxial strain. \cite{wse2_pl_strain_1, wse2_pl_strain_2, wse2_pl_strain_3, abir_wrinkled_wse2}

\begin{figure}
    \centering
    \includegraphics[width=1\linewidth]{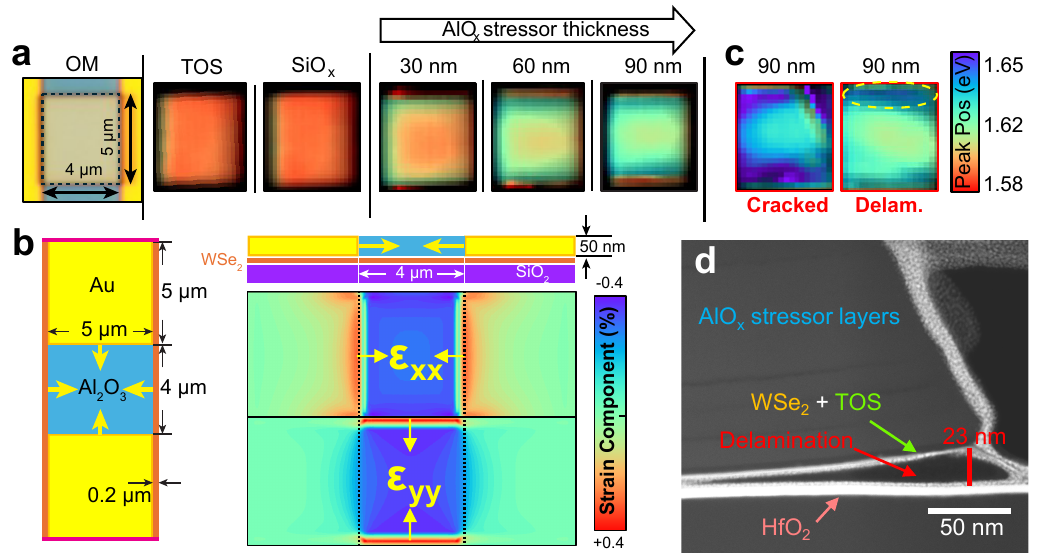}
    \caption{\textbf{Spatially resolved strain analysis.} \textbf{a,} hyperspectral maps photoluminescence peak position of one FET channel throughout doping, passivation, and 3 incremental stressor depositions. Brightness corresponds to peak intensity, and color bar hue corresponds to peak energy. \textbf{b,} top and side illustration showing the setup and dimensions of the finite-element models and maps of the simulated $\varepsilon_{xx}$ $\varepsilon_{yy}$ strain tensor components. \textbf{c,} Example PL maps of two damaged FET channels at 90 nm stressor thickness. The cracked channel shows sharp non-discontinuities, while the edge delamination shows PL amplitude dimming and peak energy inversion at the edge. \textbf{d,} Cross-section STEM image showing the edge of the channel in a damaged device, showing prominent delamination.}
    \label{fig:maps_fea}
\end{figure}

Fig.~\ref{fig:illusts_pl}e plots spatially averaged channel PL spectra taken after each processing step, from TOS doped monolayer to SiO\textsubscript{x} passivation (labeled as 0 nm stressor) to multiple {AlO\textsubscript{x}} stressor depositions. 
The PL peak does not significantly change before passivation and then steadily shifts up in energy under incremental {AlO\textsubscript{x}} thicknesses. 
Fig.~\ref{fig:illusts_pl}f tracks the spatially aggregated PL peak position statistics through stressor deposition steps. The quartile boxes show the spread of PL energies sampled, spanning the channel regions in an array of 5 separate FETs on the same chip. 
The correlation between peak energy and stressor thickness is linear, and the blueshift corresponds with the application of compression. 
The right axis of Fig.~\ref{fig:illusts_pl}f shows the corresponding calculated strain as a function of stressor thickness, assuming biaxial strain (justified in Fig.~\ref{fig:maps_fea}) and a shift rate of --120 meV/\% strain. 
At 90 nm thick, the peak shift corresponds to a maximum biaxial compression of {--0.22~\textpm~0.04\%}. \par

In Fig.~\ref{fig:maps_fea}, we assess the limits and possible failure mechanisms of our compressive stressor films. We use spatially resolved hyperspectral PL mapping to analyze the uniformity and distribution of strain in each device channel for each stressor thickness. 
Fig.~\ref{fig:maps_fea}a shows an optical image and corresponding maps of the PL peak energy of an example FET through initial TOS doping, after {SiO\textsubscript{x}}\ passivation, and as a function of stressor thickness. 
Before stressor deposition, the PL energy in the channel is uniform. As the stressor is added, the channel exhibits a global upshift, with a smaller upshift along the free edges. \par

To interpret these maps and determine the types of strain being applied, we employed finite-element analysis (FEA). Fig.~\ref{fig:maps_fea}b shows the construction of the simulation and maps of the simulated strain components $\varepsilon_{xx}$ and $\varepsilon_{yy}$. The strain profiles show good agreement with the PL maps. In the middle of the channel, the $\varepsilon_{xx}$ and $\varepsilon_{yy}$ component magnitudes are within 10\% of each other, justifying the biaxial assumption when estimating the strain-PL peak conversion rate. \cite{wse2_pl_strain_1} The tensile edges in $\varepsilon_{yy}$ created by substrate anchoring were also reproduced in FEA, corresponding to the edge redshift observed in PL mapping. For details on the FEA, see Methods and Supplementary Table S1.\par

Another key consideration in a stressed-film architecture is the maximum magnitude of in-plane strain that can be achieved before out-of-plane deformation or loss of structural integrity occurs.
Fig.~\ref{fig:maps_fea}c shows PL maps from two different FETs with 90 nm thick stressor. In these maps, we observe that some channels exhibit non-uniform PL peak distributions or dimming of PL intensity that correlate with degradation in electrical performance. 
Of the 22 initially pristine FETs, only 4 showed noticeable non-uniformity at 60 nm or less of stressor. By 90 nm, as many as 10 showed noticeable non-uniformity, suggesting that the structures are close to their in-plane strain limits. 
At 120 nm, all FETs showed non-uniformity in peak energy and intensity in conjunction with drastically degraded electrical performance. We interpret these phenomena as signs of mechanical instability, with channel fracture leading to spatially nonuniform PL peak energies and edge delamination leading to blueshifts or dimming of PL intensity. 
To verify, Fig.~\ref{fig:maps_fea}d shows the cross-sectional STEM of a free edge of a FET channel displaying a sharp blueshift in PL. In the STEM image, the stressor is bending upward, delaminating the {WSe\textsubscript{2}} from the underlying substrate, with a void undercutting several hundred nm inward from the edges. 
For reference, we note that we previously observed similar instabilities in tensile-strained {MoS\textsubscript{2}} channels at sufficient stressor thickness; though the nature and onset of the instabilities were different, as would be expected for different types of strain. \cite{yue_mos2_enh} \par

We observed that FETs exhibiting mechanical instabilities also showed non-monotonic changes in performance, exhibiting sudden and sharp decreases in current density, indicating that some form of failure had occurred. 
Thus, for systemic consistency, we only analyze the transport behavior from the 12 FETs that exhibited monotonic mobility trends and no indications of cracking or delamination failure under PL mapping. Extended Data Fig.~\ref{fig:ex_reject} contains a table of PL maps of FETs with 90 nm {AlO\textsubscript{x}}, detailing the rejection and acceptance of each for transport analysis.

\section*{Enhancing {WSe\textsubscript{2}}\ transport via compression}
\label{sec:transport}

\begin{figure}
    \centering
    \includegraphics[width=0.5\linewidth]{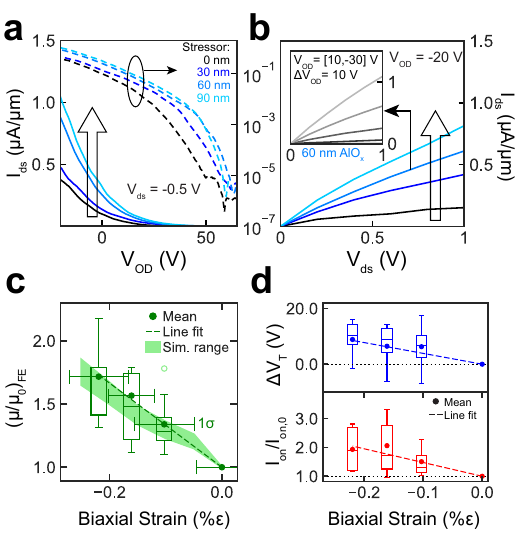}
    \caption{\textbf{Strain-enhancement of {WSe\textsubscript{2}}\ p-FET performance.} 
    \textbf{a,} Current density ($I_\text{ds}$) versus overdrive voltage ($V_\text{OD}$) transfer characteristics of a representative {WSe\textsubscript{2}}\ p-FET  for increasing stressor thicknesses (blue shades). The left and right axes show current density in linear and log scales. 
    \textbf{b,} Corresponding Current density versus drain voltage $V_{ds}$ output characteristics at constant V\textsubscript{OD} for increasing stressor thicknesses. The inset shows output characteristics for different V\textsubscript{OD} at a single 60 nm AlO\textsubscript{x} thickness. 
    \textbf{c,} Plot of measured (green points) and computed (green band) field-effect mobility factor $(\mu/\mu_0)_\text{FE}$ versus compressive (negative) biaxial strain. 
    The mobility factor tracks the change in mobility relative to the initial mobility in each p-FET. Green points show averaged measured mobility factor across 12 {WSe\textsubscript{2}}\ p-FETs, with errors representing quartile statistics in both mobility and strain uncertainty. 
    The green band represents the variability in computed mobility rising from uncertainty in the assumed experimental parameters. The mean initial mobility $\mu_\text{0,FE} =$ 3.5±1{cm\textsuperscript{2}/V\textperiodcentered s}. The mobility factor strain-tuning rate is 340{\%/\%\textgreek{e}}.
    \textbf{d,} Plots of corresponding  quartile statistics for the V\textsubscript{T} shift ($V_\text{T}-V_\text{T,0}$) and on-current factor ($I_\text{on}/I_\text{on,0}$) versus compressive biaxial strain. \mbox{V\textsubscript{T,0}~=~--30~V}, I\textsubscript{on,0}~=~0.4±0.16~μA/μm. The V\textsubscript{T} shift rate is 50 V/\%\textgreek{e} and the on-current factor strain-tuning rate is 460~{\%/\%\textgreek{e}}.}
    \label{fig:transport}
\end{figure}

In Fig.~\ref{fig:transport}, we use three-point electrical transport measurements to analyze the evolution in performance of {WSe\textsubscript{2}} p-FETs under compressive strain. 
Fig.~\ref{fig:transport}a shows the transfer characteristic evolution versus stressor thickness of a representative {WSe\textsubscript{2}}\ p-FET, with respect to overdrive voltage ($V_\text{OD}$) at \mbox{$V_\text{ds} =$--0.5~V}. 
Fig.~\ref{fig:transport}b shows the output characteristic for the same FET versus stressor thickness, with $V_\text{OD} =$ --20 V. 
Fig.~\ref{fig:transport}b inset shows the output characteristic at varying $V_\text{OD}$ for 60 nm of stressor. 
The transfer characteristics display increasing transconductance ($g_m$) with increasing stressor thickness, while the output characteristic remains linear across all stressor thicknesses and V\textsubscript{OD}. 
Extended data Fig.~\ref{fig:ex_TLM} shows transfer length measurements at each process step, which suggests that the FETs are channel-dominated and that there is no systematic change in contact resistance with strain.

For 2D FETs, small differences in material quality, environment, and process conditions disproportionately affect the contact resistance, defect density, and strain, leading to variability in the initial performance metrics. As such, conclusions made from single device measurements are often unreliable. 
We account for this unreliability by measuring the evolution in performance for 12 well-functioning FETs, selected based on the criteria discussed in Extended Data Fig. \ref{fig:ex_reject} and Supplementary Discussion 3. 
To allow direct comparison that accounts for the initial variability, we also normalize the mobility and on-current to tuning factors using the initial performance of each FET, and calculate the change in the threshold voltage.
Figures \ref{fig:transport}c-d visualize the statistical evolution of the mobility factor $(\mu/\mu_0)_\text{FE}$ at the point of peak-$g_{m}$, threshold voltage shift ($V_\text{T}-V_\text{T,0}$), and on-current factor  ($I_\text{on}/I_\text{on,0}$) at a fixed overdrive \mbox{$V_\text{OD}=$ --20 V}. See Data Fig. \ref{fig:ex_leakage} for the transfer characteristics and leakage currents of all FETs, and Supplementary Discussion 4 for details on parameter extraction. The green shaded region in Fig.~\ref{fig:transport}c shows the predicted mobility from simulations, elaborated further in the simulation section.\par 

After {SiO\textsubscript{x}} passivation (0 nm), the initial mean field-effect mobility was \mbox{$\mu_\text{0,FE} =$ {3.5~\textpm~1~cm\textsuperscript{2}/V\textperiodcentered s}}, threshold voltage was \mbox{$V_\text{T}=$ {--18~\textpm~8 V}}, and on-current was \mbox{$I_\text{on}=${1.3~\textpm~0.16~\textgreek{m}A/\textgreek{m}m}}.
For each parameter, we perform a linear fit with respect to strain to obtain the strain-tuning rates. Least importantly, the threshold voltage shifts linearly at a rate of {50~\textpm~32~V/\%\textgreek{e}}, with a maximum shift of 15 V. 
These shifts are smaller than the typical hysteresis \mbox{$V_\text{T,up} - V_\text{T,down} =$ 20 V} shown in Extended data Fig.~\ref{fig:ex_doping_pasiv}, indicating that any threshold voltage shifts are below the uncertainty in measurement. 
More importantly, we obtain strain-tuning rates of the mobility factor \mbox{$\left[\frac{\partial(\mu/\mu_0)_\text{FE}}{\partial \varepsilon}\right]$ = 340~\textpm~95 {\%/\%\textgreek{e}}} and on-current factor \mbox{$\left[\frac{\partial(I_\text{on}/I_\text{on,0})}{\partial \varepsilon}\right]$= 460~\textpm~340 {\%/\%\textgreek{e}}}.
At {--0.22~\textpm~0.04\%}\ strain, the mean mobility factor is {70\%} and the mean on-current factor is {89\%}. 
In comparison, the most up-to-date mobility factor strain-tuning rate for p-Si obtained from experimentally validated simulation is approximately 70{\%/\%\textgreek{e}}. \cite{roisin_strained_si_enh} For reference, the mobility and on-current factors achieved in historical Si p-FETs using various straining techniques varied widely from 25\% to 70\%. \cite{Hoyt_strained_si_enh, ibm_strained_si_enh, intel_strained_si_enh, strained_si_1} Overall, these results demonstrate superior strain engineering potential for {WSe\textsubscript{2}} compared to p-Si.


\section*{\textit{Ab-initio} simulation and parametric analysis of strained {WSe\textsubscript{2}}}

Next, we apply simulations to unravel the underlying mechanism behind the observed mobility enhancement, validate the simulations against the measurements, and perform predictive parametric analysis of mobility enhancement under varying strain, carrier concentration, impurity concentration, and dielectric environments. As shown in Extended Data Fig.~\ref{fig:sim_flow}, we recently developed a multi-scale computational framework that systematically integrates first-principles calculations with full-band transport modeling, to predict the strain-induced mobility enhancement in TMDs, and validated the model against the tensile strain enhancement measured in {MoS\textsubscript{2}}. \cite{taseen_tmd_strain} 
Here we apply the same model to simulate the hole carrier transport in monolayer {WSe\textsubscript{2}} under compressive strain, using experimental system parameters and materials as inputs to guide the simulation design and enhance the applicability of predictions. 

\begin{figure}
    \centering
    \includegraphics[width=1\linewidth]{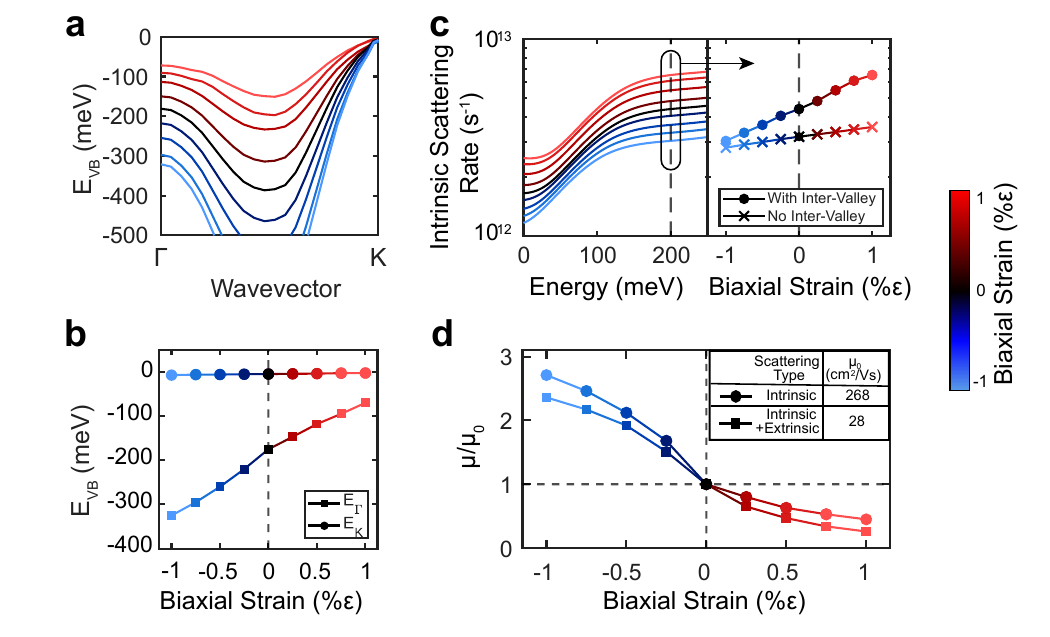}
    \caption{\textbf{Strain-induced hole mobility enhancement from modifying inter-valley scattering.}
    \textbf{a,} Computed valence band structure along the \textgreek{G}--K direction of monolayer WSe\textsubscript{2} under varying biaxial strain. Shown in the color bar on the right, in all panels, compressive, unstrained, and tensile strain regimes are represented by blue, black, and red, respectively.  
    \textbf{b,} Energy of local valence band maxima at the \textgreek{G} (square) and K (circle) valleys versus biaxial strain.  
    \textbf{c,} Left, intrinsic scattering rates versus energy under varying strain, including both intra-valley and inter-valley scattering; right, intrinsic scattering rates versus biaxial strain at an energy of 200 meV comparing the relative intrinsic scattering rate without (cross) and with (circle) inter-valley scattering allowed. \textbf{d,} Computed mobility factor versus biaxial strain, comparing the intrinsic mobility (circle) and intrinsic + extrinsic mobility (square). The inset shows the relative computed initial mobility. Fig.~\ref{fig:illusts}c--d assume $T$ = 300 K, $p = 5 \times 10^{12}\,\text{cm}^{-2}$, $n_\text{imp} = 2.5 \times 10^{12}\,\text{cm}^{-2}$, with SiO$_2$ as the dielectric environment.
}
\label{fig:illusts}
\end{figure}

Fig.~\ref{fig:illusts}a plots the valence band structure of monolayer {WSe\textsubscript{2}} at different biaxial strains, while Fig.~\ref{fig:illusts}b plots the valence band maxima (VBM) of the K and \textgreek{G} valleys versus strain. Strain is shown in the color bar from --1\% (compressive, blue) to +1\% (tensile, red). For the numerical results, see Supplementary Table S8. At zero strain, the K valley is dominant for hole transport. 
Under tensile strain, the \textgreek{G} valley maximum shifts up to converge towards the K valley maximum, while under compressive strain, the \textgreek{G} valley shifts down, separating away from the K valley maximum. This strain-modulated valley tuning is consistent with both previous simulations and optical spectroscopy studies. \cite{wiktor2016absolute, cheng2020using, shen2016strain} 
Critically, the change in valley maximum separation under strain reconfigures the fundamental electronic landscape and determines the energy-dependent scattering phase space for carriers by allowing or suppressing inter-valley scattering.  \par
The left panel of Fig.~\ref{fig:illusts}c plots the room temperature intrinsic hole scattering rate with intrinsic phonons versus energy for varying strain. The intrinsic scattering includes contributions from acoustic deformation potential (ADP), optical deformation potential (ODP), polar optical phonon (POP), inter-valley (IV), and piezoelectric (PZ) scattering mechanisms. With increasing compressive strain, the scattering rate at all energies decreases notably. To understand the dominant mechanism, the right panel of Fig.~\ref{fig:illusts}c shows the intrinsic scattering rate versus strain at 200 meV, both including and excluding IV scattering. For the numerical results, see Supplementary Table S9. At zero strain, the IV scattering dominates over other scattering mechanisms and has a much higher tuning rate versus strain. We interpret this trend as follows: Under compression, the increasing energy separation between the \textgreek{G} and K valleys will require more energy for holes to scatter from the lower \textgreek{G} valley to the higher-energy K valley. Given finite phonon energies and thermal distributions at room temperature, this larger energy barrier reduces the transition probability. Conversely, tensile strain reduces the energy barrier, enhancing IV scattering. \par

Fig.~\ref{fig:illusts}d shows the predicted hole mobility factor ($\mu/\mu_0$) versus strain $\varepsilon$, assuming only intrinsic (including IV scattering) and the combination of intrinsic + extrinsic scattering. The inset shows the predicted initial mobility $\mu_0$ for each case. These calculations are performed at $T = 300\,\text{K}$, $p = 5 \times 10^{12}\,\text{cm}^{-2}$, $n_\text{imp} = 2.5 \times 10^{12}\,\text{cm}^{-2}$, with SiO$_2$ as the dielectric environment. 
The intrinsic mobility (circle) accounts only for ADP, ODP, POP, IV, and PZ scattering, while the extrinsic mobility (square) also includes charge impurity (CI) scattering and surface optical (SO) phonon interactions. In both cases, the mobility is enhanced under compressive strain and suppressed under tensile strain. For the numerical results, see Supplementary Table S10. Next, we validate the simulation against the experimental data. 
Shown in Fig.~\ref{fig:transport}c, the green band overlay shows the predicted enhancement rate, assuming a range in impurity and carrier densities that reflects the experimental variability. The theory and experiment agree well, indicating that the computed enhancement rates correctly capture the underlying mechanisms.\par

\begin{figure}[htp]
\centering
\includegraphics[width=0.97\linewidth]{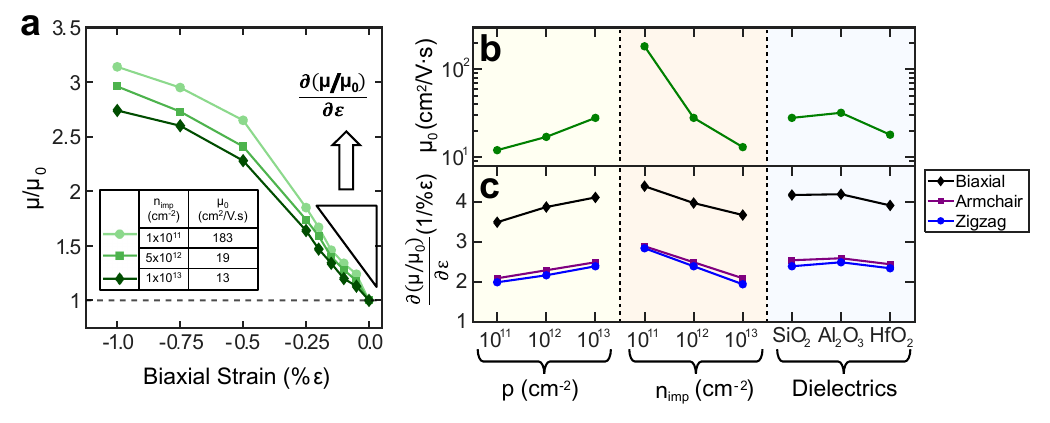}
\caption{\textbf{Parametric analysis of strain-modulated hole mobility enhancement in {WSe\textsubscript{2}}.}
\textbf{a,} Mobility factor versus biaxial compressive strain for different impurity concentrations. \textbf{b-c,} Parametric analysis versus carrier concentration $(p)$, impurity concentration $(n_\mathrm{imp})$, and different dielectric environments of \textbf{b,} unstrained initial hole mobility $\mu_0$ (green) and \textbf{c,} mobility factor strain-tuning rate {$\frac{\partial(\mu/\mu_0)}{\partial \varepsilon}$}\ (1/\%\textgreek{e}) and evaluated at --0.25\% compressive strain. Color represents different strain types: biaxial (black), uniaxial armchair (purple), and uniaxial zigzag (blue). In all plots, unless specifically being varied, analyses assume $T$ = 300 K, $p =  10^{13}$\,cm$^{-2}$, $n_{\text{imp}} = 2.5 \times 10^{12}$\,cm$^{-2}$, and a SiO$_2$ dielectric.}
\label{fig:sim_param_vary}
\end{figure}

Notably, while the initial mobility is an order of magnitude different between the intrinsic and extrinsic cases, the mobility factor strain-tuning rate {$\left[\frac{\partial(\mu/\mu_0)}{\partial \varepsilon}\right]$}\ is very similar. This surprising observation begs the question of how robust the rate is. 
We thus perform parametric analysis to examine how variations in material and environmental parameters influence the rate of mobility enhancement under strain.  Fig.~\ref{fig:sim_param_vary}a shows the computed hole mobility factor versus compressive strain at different impurity densities ($n_{\mathrm{imp}} = \{10^{11}, 5\times10^{12}, 10^{13}\}\,\mathrm{cm}^{-2}$) shown as light to dark green. These impurity densities capture the range of values commonly observed in experiments. We keep other parameters fixed with $T$ = 300 K, $p = 10^{13}\,\text{cm}^{-2}$, and SiO$_2$ as the dielectric environment. The inset shows the corresponding initial mobilities $\mu_0$ for each impurity density. For the numerical results, see Supplementary Table S11. \par

To explicitly quantify the sensitivity of the mobility enhancement to critical parameters, Figs.~\ref{fig:sim_param_vary}b-c respectively plot changes in the unstrained mobility ($\mu_0$) and the strain-tuning rates {$\left[\frac{\partial(\mu/\mu_0)}{\partial \varepsilon}\right]$}\ versus carrier density ($p = \{10^{11}, 10^{12}, 10^{13}\}\,\mathrm{cm}^{-2}$), 
impurity concentration ($n_{\mathrm{imp}} = \{10^{11}, 10^{12}, 10^{13}\}\,\mathrm{cm}^{-2}$), 
and dielectric environment (SiO$_2$, Al$_2$O$_3$, and HfO$_2$). We further compare the relative tuning rates for biaxial (black), uniaxial armchair (purple), and uniaxial zig-zag (blue), with the slopes extracted at --0.25\% compression. Unless specifically varied, all other parameters are kept fixed at $T = 300\,\text{K}$, $p = 10^{13}$\,cm$^{-2}$, $n_{\text{imp}} = 2.5 \times 10^{12}$\,cm$^{-2}$, with SiO$_2$ as the dielectric environment. For the numerical results, see Supplementary Tables S12-13. \par

The parametric analysis yields three important insights. First, the enhancement rate from biaxial strain is roughly twice the enhancement rate of uniaxial strain. Armchair produces a slightly higher enhancement than zig-zag, but the difference is negligible. 
This effect can easily be understood as biaxial strain tunes the band positions at roughly twice the rate of uniaxial strain. Second, across two orders of magnitude change in critical parameters, the computed initial mobility changes by an order of magnitude, ranging from $\mu_0$ = 10--150 {cm\textsuperscript{2}/V\textperiodcentered s}. This variation in initial mobility is well understood as a competition between changes in impurity-induced scattering, electrostatic screening, and phonon scattering. \cite{leveillee2023ab, zhang2022phonon} Yet, over two orders of magnitude variation in carrier concentration, impurity density, and dielectric environment, the mobility enhancement rate is robust, ranging from 200--300\%/\%\textgreek{e} (uniaxial) and 350--400\%/\%\textgreek{e} (biaxial). The enhancement's robustness originates from strain-suppressing IV scattering through a direct reconfiguration of the band structure. This represents a multiplicative effect on the electronic phase space, fundamentally distinct from the additive scattering rates governed by Matthiessen's rule.

The important implication is that strain engineering leads to a multiplicative enhancement in performance when combined with other mobility engineering approaches like modulating the dielectric environment and controlling doping or defects.

\section*{Conclusions}
There is an urgent need to understand the interfacial mechanics that determine how CMOS processes impact device performance in 2D material beyond-silicon electronics. This study provides a first demonstration of how compressive strain enhances the electrical performance of p-type monolayer {WSe\textsubscript{2}}\ FETs. The stressed film-based technique is lithography-compatible and applies uniform, localized compression to the 2D FET channels. Simulations show that the enhancement observed arises from the suppression of inter-valley scattering of holes in {WSe\textsubscript{2}}\ under compression. The strain-induced mobility enhancement rates exceed those of strained complementary silicon by 4-10x, and are effectively independent of the changes in mobility offered by engineering carrier density, dielectric environment, and impurity density. These results underscore the multiplicative nature of strain-engineering as an important strategy to improve hole transport in {WSe\textsubscript{2}}. Integrating strain into 2D electronics also unlocks new material physics and device concepts that actively utilize strain, such as Moire engineering, \cite{pena_moire} slidetronics, \cite{slidetronics} or solid state optics based on pseudo-magnetic fields. \cite{stressor_engineering} 
In combination with recent results on the role of tensile strain to enhance n-type transistors,\cite{yue_mos2_enh, yang_mos2_enh, shin_mos2_enh2, pop_mos2_enh, chen_mos2_enh}, these results incentivize the use of strain engineering in future high-performance TMD transistor node designs and new computational paradigms.

\section*{Methods}\label{sec11}

\subsection*{Gold-assisted exfoliation of TMDs}
We adapted published methods for gold-assisted large-area exfoliation of monolayer and bilayer TMDs. \cite{gold_exfol_nature} We used E-beam evaporation (0.5 \AA/s) to deposit a 30 nm Au film on a fresh, clean (100) Si wafer. Then, we spin-coated (at 3000 RPM for 60 s) a few-micron layer of PPC using a 15\% wt. anisole solution to act as a supporting film on top of the Au. 
For the exfoliation, we picked up a piece of the Au/PPC film from the wafer using a Kapton frame and attached the bottom surface of the Au film to a bulk {WSe\textsubscript{2}} crystal, supplied by HQ Graphene. To improve contact, we heated the structure to 180 \textdegree C for 60 s on a hotplate. After cooling, we used tweezers to carefully release the bulk crystal from the Au film, completing the exfoliation of the WSe\textsubscript{2} onto Au.
Finally, we dissolved the PPC using three methanol rinses and etched the gold film using KI from Sigma Aldrich. Supplementary Fig.~S1 shows the {WSe\textsubscript{2}} surface topography throughout the fabrication stages. This approach produces mm-scale monolayers on the Au film, with 100 $\mu$m scale bilayer patches. 

\subsection*{Fabrication of p-doped {WSe\textsubscript{2}} FETs}
Fig.~\ref{fig:ex_fab} illustrates the fabrication flow of p-doped {WSe\textsubscript{2}}\ FETs, including the incremental stressor deposition cycle. 
We selected bilayer {WSe\textsubscript{2}} patches obtained from gold-assisted exfoliation, and patterned arrays of multiple FETs on the same bilayer patch. This allowed us to minimize uncertainties around sample-to-sample variation, and material quality. On each bilayer, we started by using e-beam lithography (EBL) to pattern the channels. Pure XeF\textsubscript{2} gas (at 3 Torr, for 120 s) was used to etch any excess WSe\textsubscript{2}. All EBL lithography steps were performed with 200 nm thick 495PMMA A4 resist and using the Raith EBPG5150 system at 100 kV, 330 \textgreek{m}C/cm\textsuperscript{2} dosage. We used 0 \textdegree C MIBK developer (1 part MIBK, 3 parts IPA) to develop PMMA resist; development time was variable. 
After stripping the EBL resist using acetone, we used a custom oven to perform a vacuum anneal for 3 h at 200 \textdegree C, 10\textsuperscript{-7} Torr. This step reduced the occurrence of interfacial bubbles and surface adsorbates. We used a Tergeo oxygen plasma cleaner and exposed the bilayer {WSe\textsubscript{2}}\ channel to remote oxygen plasma at 50 W for 3 min. This step induces a self-limiting reaction that oxidizes the top layer and leaves a heterostructure of 1L {WSe\textsubscript{2}}\ with approximately 2 nm of tungsten oxyselenide({WO\textsubscript{x}Se\textsubscript{y}}, Labeled TOS), a p-dopant, on top. 
The TOS both reduces contact resistance and ensures heavy p-doping in the channel through the device processing.\cite{sihan_TOS, borah_TOS_doping} We then used EBL to pattern electrodes and E-beam evaporation to deposit 5 nm Pd, 35 nm Au contacts with Pd deposited at 0.1 \AA/s and Au deposited at 0.5 \AA/s. 
After this step, the p-FETs were fully functional, with channel dimensions of 4 $\mu$m long by 5 $\mu$m wide. We performed one final EBL step to create PMMA windows for passivation film and stressor deposition with dimensions of 5 $\mu$m long by 4.6 $\mu$m wide (with respect to the direction of current flow). This way, the window overlapped the metal contacts by 500 nm, but was narrower than the channel by 200 nm on each side. 
The lateral recess of the stressor maximized compression of the WSe\textsubscript{2} by preventing anchorage to the substrate, and the longitudinal overlap ensured that the channel strain was uniform along the direction of current flow.
Finally, we used e-beam evaporation to deposit a 6 nm passivation layer of {SiO\textsubscript{x}} on the channel using the window. We chose a slow rate of 0.05 \AA /s to minimize damage to the delicate TOS layer and maintain the heavy p-doping.

\subsection*{Stressor deposition}
We deposited the compressive {AlO\textsubscript{x}}\ using E-beam evaporation at room temperature, with a base pressure of 10\textsuperscript{-7} Torr. The evaporation rate of the Al\textsubscript{2}O\textsubscript{3} source was 0.5 \AA /s. We deposited {AlO\textsubscript{x}}\ at no more than 30 nm per session to limit substrate heating and to allow systematic measurements of the change in stress. Through the whole process, we left the PMMA window in place without liftoff, so the film only interacts with the channel, and to minimize uncertainties due to interfacial residue or layer misalignment that would be caused by repeated lithography steps for each layer. 

\subsection*{Thin film stress extraction}
We used standard wafer curvature measurements to quantify the thin film stress of the AlO\textsubscript{x} film. Then we used a Dektak DXT-A stylus profilometer to measure the wafer deflection and effective radius of curvature of a fused silica cover slip with a known thickness of 150 \textgreek{m}m before and after AlO\textsubscript{x} deposition with two different thicknesses, 196.6 \textpm 0.07 nm and 311.88 \textpm 0.15 nm. We assessed the film thickness using a separate, partially shadow-masked cover slip in the same deposition and measured the step height in the same profilometer. We then used Stoney's equation to extract the stress from the change in curvature.\par

Fig.~\ref{fig:ex_wafer_bend} shows the topography profiles before and after stressor deposition, and provides the extracted stress and thin film force values for two film thicknesses.  

\subsection*{Photoluminescence mapping}
We performed PL measurements using a Nanophoton Confocal Raman system with a laser wavelength of 531.94 nm, \mbox{600 line/mm} grating, and NA 0.9 100\texttimes\ objective. For hyperspectral mapping, we used the horizontal row imaging mode at \mbox{0.3 mW/line} at a spatial resolution of \mbox{200 nm/pixel}. The pixel aspect ratio was square, and each row was integrated over 3 s. 

\subsection*{Electrical measurements}
We performed electrical measurements using an atmospheric, room-temperature probe station connected to a Keithley 4200A-SCS parameter analyzer. We found that there is no problem with having probes punch through the deposited layers to make electrical contacts to the source and drain, and used silver paste to make contact the the global silicon back gate. We made gate sweeps in voltage increments of 0.5 V at 3 increments per second. For transfer characteristics, we swept V\textsubscript{gs} from +40 V to --50 V then back to +40 V, over the course of approximately 2 minutes. For output characteristics, we swept V\textsubscript{ds}
from --1 V to 1 V over the course of approximately 2 s, incrementing V\textsubscript{gs} in steps of 10 V between each sweep. We chose this low V\textsubscript{ds} range to minimize current-related thermal wear over the many measurement iterations. The numerical data of transport parameters can be found in Supplementary Tables~S2-S7.

\subsection*{{WSe\textsubscript{2}}\ transport simulation framework}

As shown in Fig.~\ref{fig:sim_flow}, we investigated the carrier transport in monolayer {WSe\textsubscript{2}} under compressive strain through a multi-scale computational framework that systematically integrates first-principles calculations with full-band transport modeling. The workflow consists of three major stages: (i) first-principles determination of electronic and vibrational properties, (ii) computation of energy-dependent scattering rates from intrinsic and extrinsic mechanisms, and (iii) evaluation of the strain-modulated hole mobility.\cite{taseen_tmd_strain} 

We used density functional theory to calculate the electronic and phononic properties as implemented in Quantum ESPRESSO~\cite{giannozzi2017advanced}, employing the Perdew-Burke-Ernzerhof (PBE) functional within the GGA~\cite{perdew1996generalized} and PAW pseudopotentials.~\cite{giustino2017electron} We introduced a vacuum spacing of 20~\AA\ to avoid non-physical coupling between periodic images. We used plane-wave cutoffs of 80~Ry and 320~Ry for wavefunctions and charge densities, respectively, with a $24 \times 24 \times 1$ Monkhorst-Pack $k$-grid for relaxations and electronic structure. We obtained phonon spectra via DFPT~\cite{baroni2001phonons} using a $48 \times 48 \times 1$ $q$-grid. 

We evaluated the carrier mobility within the linear-response regime using the Kubo--Greenwood formalism \cite{lundstrom2002fundamentals}, which provides a fundamental link between microscopic carrier dynamics and macroscopic transport phenomena. In this framework, the key quantity governing charge transport is the energy-dependent momentum relaxation time, which encapsulates the cumulative effects of the intrinsic and extrinsic scattering mechanisms on carrier motion. The relaxation time was obtained using Fermi’s golden rule (FGR), following the detailed derivation.\cite{taseen_tmd_strain}

By systematically incorporating strain-induced changes in both the electronic band structure and phonon spectra, we quantitatively analyzed their individual and combined impacts on scattering times and carrier mobility. All parameters were calculated at each value of applied strain for {WSe\textsubscript{2}}. We obtained the essential physical parameters from the first-principles calculations, including group velocity, density of states, deformation potentials, elastic constants, phonon energies, sound velocity, and mass density. \cite{taseen_tmd_strain}

\subsection*{Finite-element strain simulations}

We used the commercially available ABAQUS Standard FEA solver to model and simulate the strain state in the channel region under stressors and Au contacts by using previously developed traction separation relations to describe the strain transfer.\cite{yue_stressor} Material properties and dimensions of different parts are summarized in Supplementary Table S1. 
We modeled the interfacial properties between the layers (e.g., {WSe\textsubscript{2}}, Au, {SiO\textsubscript{x}}\ and {AlO\textsubscript{x}}) as surface-to-surface interaction using cohesive behavior with specified damage evolution properties. We used a trapezoidal traction--separation law with an interfacial traction coefficient $K_b$, a damage initiation threshold $\delta_d$, and a maximum plastic deformation $\delta_m$ to model the van der Waals (vdW) interaction between the SiO\textsubscript{2} substrate and the monolayer WSe\textsubscript{2}. We defined another interfacial traction coefficient $K_t$ ($K_t \gg K_b$) to represent the stiffness of the top surface interactions between the van der Waals (vdW) material and the deposited thin film stressor. 
We used the previously published relation between the strain decay length $\lambda$ and the interfacial traction coefficient $K_b$ to estimate the value of $K_b$ in our model. \cite{yue_stressor} Numerically, we used $K_t = 10{,}000~\text{MPa/m}$ and $K_b = 10~\text{MPa/m}$. 
In this model, the mesh size is determined by the smallest dimension, in this case, the thickness of the 2D monolayer WSe\textsubscript{2}, which we assume to be 0.001~\textgreek{m}m. We used C3D8R elements with a mesh size ranging from 0.06--0.12~\textgreek{m}m. The built-in stress of {AlO\textsubscript{x}} is defined as a pre-defined stress field of 350.0 MPa. For solver stability, we implemented a tie constraint the lateral surfaces of Au contacts and {AlO\textsubscript{x}} stressor and fixed (both translation and rotation) boundary conditions to all the bottom nodes of the {SiO\textsubscript{x}} substrate.

\subsection*{Cross-section scanning-transmission electron microscopy}

Cross section samples were prepared following standard FIB lift-out and thinning procedures in a Thermo Fisher Scientific Helios 600i DualBeam FIB-SEM. Initial milling was performed at 30 kV, with final milling and cleaning performed at 5 kV and 2 kV to minimize sample damage. We acquired electron microscopy data on an aberration-corrected Thermo Fisher Scientific Themis Z (S)TEM operated at 80 kV with a standard double tilt holder. ADF-STEM images were collected with a 25.2 mrad convergence semi-angle, a 115 mm camera length, a 2 \textgreek{m}s/pixel dwell time, and a probe current of 20--30 pA. 

\section*{Data availability}

The raw data used for analysis in this work have been archived with open access at the Illinois Data Bank: \url{https://databank.illinois.edu/datasets/IDB-6891048?code=ykgI13wipT4qNf1TMfU1CwVlA5YlrEmCSU-H16AgkGk}

\section*{Acknowledgments}

This work was primarily supported by the National Science Foundation (NSF) through the Center for Advanced Semiconductor Chips with Accelerated Performance (ASAP) Industry-University Cooperative Research Center under NSF Cooperative Agreement No. EEC-2231625. The authors acknowledge partial support by the Illinois Materials Research Science and Engineering Center (I-MRSEC) under Award DMR-2309037. The authors acknowledge the use of facilities and instrumentation supported by NSF through the University of Illinois Materials Research Science and Engineering Center DMR-2309037. This work was performed in the Holonyak Micro and Nanotechnology Laboratory (HMNTL) and the Materials Research Lab (MRL) at the University of Illinois.

\section*{Contributions}
A.M.v.d.Z., Y.Z., and H.L.Z. conceived the project. H.L.Z. performed the fabrication, characterization, and analysis of the strained {WSe\textsubscript{2}} FETs. S.M.T.S.A. carried out first-principles calculations and developed the multi-scale transport framework. D.Y. performed mechanical experimentation on {AlO\textsubscript{x}} films. Z.I. performed the finite-element analysis. Z.M. performed cross-sectional STEM imaging under the supervision of P.Y.H., Y.Z., D.Y., and S.C. provided experimental design inputs. S.R. guided the simulations and provided theoretical insights. A.M.v.d.Z. guided the direction of the research, experimental design, and research strategy. All authors contributed to the manuscript.

\section*{Competing interests}
The authors declare no competing interests.


\begin{thebibliography}{10}
\expandafter\ifx\csname url\endcsname\relax
  \def\url#1{\texttt{#1}}\fi
\expandafter\ifx\csname urlprefix\endcsname\relax\def\urlprefix{URL }\fi
\providecommand{\bibinfo}[2]{#2}
\providecommand{\eprint}[2][]{\url{#2}}

\bibitem{strained_si_1}
\bibinfo{author}{Thompson, S.} \emph{et~al.}
\newblock \bibinfo{title}{A 90-nm logic technology featuring strained-silicon}.
\newblock \emph{\bibinfo{journal}{IEEE Trans. Electron Devices}} \textbf{\bibinfo{volume}{51}}, \bibinfo{pages}{1790--1797} (\bibinfo{year}{2004}).

\bibitem{strained_si_2}
\bibinfo{author}{Welser, J.}
\newblock \bibinfo{title}{{NMOS} and {PMOS} transistors fabricated in strain silicon/relaxed silicon-germanium structures}.
\newblock In \emph{\bibinfo{booktitle}{1992 IEDM}}, \bibinfo{pages}{1000--1002} (\bibinfo{year}{1992}).

\bibitem{strained_si_3}
\bibinfo{author}{Kuhn, K.~J.}, \bibinfo{author}{Murthy, A.}, \bibinfo{author}{Kotlyar, R.} \& \bibinfo{author}{Kuhn, M.}
\newblock \bibinfo{title}{Past, present and future: {SiGe} and {CMOS} transistor scaling}.
\newblock \emph{\bibinfo{journal}{ECS Trans.}} \textbf{\bibinfo{volume}{33}}, \bibinfo{pages}{3} (\bibinfo{year}{2010}).

\bibitem{2d_scaling_projection}
\bibinfo{author}{Akinwande, D.} \emph{et~al.}
\newblock \bibinfo{title}{Graphene and two-dimensional materials for silicon technology}.
\newblock \emph{\bibinfo{journal}{Nature}} \textbf{\bibinfo{volume}{573}}, \bibinfo{pages}{507--518} (\bibinfo{year}{2019}).

\bibitem{tsmc_first_2dgaa}
\bibinfo{author}{Chung, Y.-Y.} \emph{et~al.}
\newblock \bibinfo{title}{First demonstration of {GAA} monolayer {MoS\textsubscript{2}} nanosheet {nFET} with {410μA/μm} {I\textsubscript{D}}, {1V} {V\textsubscript{D}} at 40nm gate length}.
\newblock In \emph{\bibinfo{booktitle}{2022 IEDM}}, \bibinfo{pages}{34.5.1--34.5.4} (\bibinfo{year}{2022}).

\bibitem{imec_roadmap}
\bibinfo{author}{de~la Rosa, C. J.~L.} \& \bibinfo{author}{Kar, G.~S.}
\newblock \bibinfo{title}{{2D}-material based devices in the logic scaling roadmap | imec}.
\newblock \bibinfo{type}{Tech. Rep.}, \bibinfo{institution}{IMEC International} (\bibinfo{year}{2025}).
\newblock \urlprefix\url{https://www.imec-int.com/en/articles/introducing-2d-material-based-devices-logic-scaling-roadmap}.

\bibitem{tsmc_2d_cfet}
\bibinfo{author}{Chung, Y.~Y.} \emph{et~al.}
\newblock \bibinfo{title}{Stacked channel transistors with {2D} materials: an integration perspective}.
\newblock In \emph{\bibinfo{booktitle}{2024 IEDM}}, \bibinfo{pages}{1--4} (\bibinfo{publisher}{Institute of Electrical and Electronics Engineers Inc.}, \bibinfo{year}{2024}).

\bibitem{tsmc_stacked_2dgaa}
\bibinfo{author}{Chung, Y.-Y.} \emph{et~al.}
\newblock \bibinfo{title}{Monolayer-{MoS\textsubscript{2}} stacked nanosheet channel with c-type metal contact}.
\newblock In \emph{\bibinfo{booktitle}{2023 IEDM}}, \bibinfo{pages}{1--4} (\bibinfo{year}{2023}).

\bibitem{2d_fet_performance_review}
\bibinfo{author}{Zeng, S.}, \bibinfo{author}{Liu, C.} \& \bibinfo{author}{Zhou, P.}
\newblock \bibinfo{title}{Transistor engineering based on {2D} materials in the post-silicon era}.
\newblock \emph{\bibinfo{journal}{Nat. Rev. Electr. Eng.}} \textbf{\bibinfo{volume}{1}}, \bibinfo{pages}{335--348} (\bibinfo{year}{2024}).

\bibitem{2d_fet_mobility_var}
\bibinfo{author}{Sim, S.} \emph{et~al.}
\newblock \bibinfo{title}{Strategic mobility engineering in {2D} semiconductor-based {FET}s for enhanced electronic devices}.
\newblock \emph{\bibinfo{journal}{Adv. Sci.}} \textbf{\bibinfo{volume}{12}}, \bibinfo{pages}{e09170} (\bibinfo{year}{2025}).

\bibitem{pfet_challenges}
\bibinfo{author}{Jiang, J.} \emph{et~al.}
\newblock \bibinfo{title}{Advancing 2d cmos electronics with high-performance p-type transistors}.
\newblock \emph{\bibinfo{journal}{Nature Communications 2025 16:1}} \textbf{\bibinfo{volume}{16}}, \bibinfo{pages}{10233--} (\bibinfo{year}{2025}).
\newblock \urlprefix\url{https://www.nature.com/articles/s41467-025-66263-0}.

\bibitem{wang_wse2_contact}
\bibinfo{author}{Wang, Y.} \emph{et~al.}
\newblock \bibinfo{title}{Does p-type ohmic contact exist in wse2–metal interfaces?}
\newblock \emph{\bibinfo{journal}{Nanoscale}} \textbf{\bibinfo{volume}{8}}, \bibinfo{pages}{1179--1191} (\bibinfo{year}{2015}).

\bibitem{al2o3_ndope}
\bibinfo{author}{Kim, S.~Y.}, \bibinfo{author}{Park, S.} \& \bibinfo{author}{Choi, W.}
\newblock \bibinfo{title}{Enhanced carrier mobility of multilayer {MoS\textsubscript{2}} thin-film transistors by {Al\textsubscript{2}O\textsubscript{3}} encapsulation}.
\newblock \emph{\bibinfo{journal}{Appl. Phys. Lett.}} \textbf{\bibinfo{volume}{109}}, \bibinfo{pages}{45} (\bibinfo{year}{2016}).

\bibitem{vacancy_doping}
\bibinfo{author}{Zhang, X.} \emph{et~al.}
\newblock \bibinfo{title}{Single-atom vacancy doping in two-dimensional transition metal dichalcogenides}.
\newblock \emph{\bibinfo{journal}{Accounts of Materials Research}} \textbf{\bibinfo{volume}{2}}, \bibinfo{pages}{655--668} (\bibinfo{year}{2021}).

\bibitem{mos2_strain_review}
\bibinfo{author}{Katiyar, A.~K.} \& \bibinfo{author}{Ahn, J.~H.}
\newblock \bibinfo{title}{Strain-engineered {2D} materials: Challenges, opportunities, and future perspectives}.
\newblock \emph{\bibinfo{journal}{Small Methods}} \textbf{\bibinfo{volume}{9}}, \bibinfo{pages}{2401404} (\bibinfo{year}{2024}).

\bibitem{shin_mos2_enh2}
\bibinfo{author}{Shin, H.} \emph{et~al.}
\newblock \bibinfo{title}{Nonconventional strain engineering for uniform biaxial tensile strain in {MoS\textsubscript{2}} thin film transistors}.
\newblock \emph{\bibinfo{journal}{ACS Nano}} \textbf{\bibinfo{volume}{18}}, \bibinfo{pages}{4414--4423} (\bibinfo{year}{2024}).

\bibitem{yang_mos2_enh}
\bibinfo{author}{Yang, J.~A.} \emph{et~al.}
\newblock \bibinfo{title}{Biaxial tensile strain enhances electron mobility of monolayer transition metal dichalcogenides}.
\newblock \emph{\bibinfo{journal}{ACS Nano}} \textbf{\bibinfo{volume}{18}}, \bibinfo{pages}{18151--18159} (\bibinfo{year}{2024}).

\bibitem{pop_mos2_enh}
\bibinfo{author}{Jaikissoon, M.} \emph{et~al.}
\newblock \bibinfo{title}{{CMOS}-compatible strain engineering for high-performance monolayer semiconductor transistors}.
\newblock \emph{\bibinfo{journal}{Nat. Electron.}} \textbf{\bibinfo{volume}{7}}, \bibinfo{pages}{885--891} (\bibinfo{year}{2024}).

\bibitem{roisin_strained_si_enh}
\bibinfo{author}{Roisin, N.} \emph{et~al.}
\newblock \bibinfo{title}{Phonon-limited mobility for electrons and holes in highly-strained silicon}.
\newblock \emph{\bibinfo{journal}{npj Comput. Mater.}} \textbf{\bibinfo{volume}{10}}, \bibinfo{pages}{1--11} (\bibinfo{year}{2024}).

\bibitem{yue_mos2_enh}
\bibinfo{author}{Zhang, Y.}, \bibinfo{author}{Zhao, H.~L.}, \bibinfo{author}{Huang, S.}, \bibinfo{author}{Hossain, M.~A.} \& \bibinfo{author}{van~der Zande, A.~M.}
\newblock \bibinfo{title}{Enhancing carrier mobility in monolayer {MoS\textsubscript{2}} transistors with process-induced strain}.
\newblock \emph{\bibinfo{journal}{ACS Nano}} \textbf{\bibinfo{volume}{18}}, \bibinfo{pages}{12377--12385} (\bibinfo{year}{2024}).

\bibitem{pena_mos2_strain}
\bibinfo{author}{Pe{\~n}a, T.} \emph{et~al.}
\newblock \bibinfo{title}{Strain engineering {2D} {MoS\textsubscript{2}} with thin film stress capping layers}.
\newblock \emph{\bibinfo{journal}{2D Mater.}} \textbf{\bibinfo{volume}{8}}, \bibinfo{pages}{045001} (\bibinfo{year}{2021}).

\bibitem{azizi_mos2_strain}
\bibinfo{author}{Azizimanesh, A.}, \bibinfo{author}{Pe{\~n}a, T.}, \bibinfo{author}{Sewaket, A.}, \bibinfo{author}{Hou, W.} \& \bibinfo{author}{Wu, S.~M.}
\newblock \bibinfo{title}{Uniaxial and biaxial strain engineering in {2D} {MoS\textsubscript{2}} with lithographically patterned thin film stressors}.
\newblock \emph{\bibinfo{journal}{Appl. Phys. Lett.}} \textbf{\bibinfo{volume}{118}}, \bibinfo{pages}{213104} (\bibinfo{year}{2021}).

\bibitem{azizi_graphene_strain}
\bibinfo{author}{Azizimanesh, A.} \emph{et~al.}
\newblock \bibinfo{title}{Strain engineering in {2D} {hBN} and graphene with evaporated thin film stressors}.
\newblock \emph{\bibinfo{journal}{Appl. Phys. Lett.}} \textbf{\bibinfo{volume}{123}}, \bibinfo{pages}{43504} (\bibinfo{year}{2023}).

\bibitem{yue_stressor}
\bibinfo{author}{Zhang, Y.} \emph{et~al.}
\newblock \bibinfo{title}{Patternable process-induced strain in {2D} monolayers and heterobilayers}.
\newblock \emph{\bibinfo{journal}{ACS Nano}} \textbf{\bibinfo{volume}{18}}, \bibinfo{pages}{4205--4215} (\bibinfo{year}{2024}).

\bibitem{sihan_TOS}
\bibinfo{author}{Chen, S.}, \bibinfo{author}{Zhang, Y.}, \bibinfo{author}{King, W.~P.}, \bibinfo{author}{Bashir, R.} \& \bibinfo{author}{van~der Zande, A.~M.}
\newblock \bibinfo{title}{Extension doping with low-resistance contacts for p-type monolayer {WSe\textsubscript{2}} field-effect transistors}.
\newblock \emph{\bibinfo{journal}{Adv. Electron. Mater.}} \bibinfo{pages}{2400843} (\bibinfo{year}{2024}).

\bibitem{oberoi_TOS}
\bibinfo{author}{Oberoi, A.} \emph{et~al.}
\newblock \bibinfo{title}{Toward high-performance p-type two-dimensional field effect transistors: Contact engineering, scaling, and doping}.
\newblock \emph{\bibinfo{journal}{ACS Nano}} \textbf{\bibinfo{volume}{17}}, \bibinfo{pages}{19709--19723} (\bibinfo{year}{2023}).

\bibitem{borah_TOS_doping}
\bibinfo{author}{Borah, A.}, \bibinfo{author}{Nipane, A.}, \bibinfo{author}{Choi, M.~S.}, \bibinfo{author}{Hone, J.} \& \bibinfo{author}{Teherani, J.~T.}
\newblock \bibinfo{title}{Low-resistance p-type ohmic contacts to ultrathin {WSe\textsubscript{2}} by using a monolayer dopant}.
\newblock \emph{\bibinfo{journal}{ACS Appl. Electron. Mater.}} \textbf{\bibinfo{volume}{3}}, \bibinfo{pages}{2941--2947} (\bibinfo{year}{2021}).

\bibitem{wse2_pl_strain_1}
\bibinfo{author}{Henríquez-Guerra, E.} \emph{et~al.}
\newblock \bibinfo{title}{Large biaxial compressive strain tuning of neutral and charged excitons in single-layer transition metal dichalcogenides}.
\newblock \emph{\bibinfo{journal}{ACS Appl. Mater. Interfaces}} \textbf{\bibinfo{volume}{15}}, \bibinfo{pages}{57369--57378} (\bibinfo{year}{2023}).

\bibitem{wse2_pl_strain_2}
\bibinfo{author}{Roy, S.}, \bibinfo{author}{Gao, J.}, \bibinfo{author}{Gao, J.}, \bibinfo{author}{Yang, X.} \& \bibinfo{author}{Yang, X.}
\newblock \bibinfo{title}{Upconversion photoluminescence of monolayer {WSe\textsubscript{2}} with biaxial strain tuning}.
\newblock \emph{\bibinfo{journal}{Opt. Express}} \textbf{\bibinfo{volume}{32}}, \bibinfo{pages}{3308--3315} (\bibinfo{year}{2024}).

\bibitem{wse2_pl_strain_3}
\bibinfo{author}{Lv, Y.} \emph{et~al.}
\newblock \bibinfo{title}{Strain-dependent optical properties of monolayer {WSe\textsubscript{2}}}.
\newblock \emph{\bibinfo{journal}{J. Phys. Chem. C}} \textbf{\bibinfo{volume}{127}}, \bibinfo{pages}{22682--22691} (\bibinfo{year}{2023}).

\bibitem{abir_wrinkled_wse2}
\bibinfo{author}{Hossein, M.~A.}, \bibinfo{author}{Zhang, Y.} \& \bibinfo{author}{van~der Zande, A.}
\newblock \bibinfo{title}{Strain engineering photonic properties in monolayer semiconductors through mechanically-reconfigurable wrinkling}.
\newblock In \bibinfo{editor}{Nielsen, C.} \& \bibinfo{editor}{Congreve, D.} (eds.) \emph{\bibinfo{booktitle}{Physical Chemistry of Semiconductor Materials and Interfaces XIX}}, vol. \bibinfo{volume}{11464}, \bibinfo{pages}{1146404} (\bibinfo{publisher}{SPIE}, \bibinfo{year}{2020}).

\bibitem{Hoyt_strained_si_enh}
\bibinfo{author}{Hoyt, J.~L.} \emph{et~al.}
\newblock \bibinfo{title}{Strained silicon {MOSFET} technology}.
\newblock In \emph{\bibinfo{booktitle}{2002 IEDM}}, \bibinfo{pages}{23--26} (\bibinfo{year}{2002}).

\bibitem{ibm_strained_si_enh}
\bibinfo{author}{Chan, V.} \emph{et~al.}
\newblock \bibinfo{title}{Strain for {CMOS} performance improvement}.
\newblock In \emph{\bibinfo{booktitle}{Proceedings of the IEEE: Custom Integrated Circuits Conference}}, \bibinfo{pages}{667--674} (\bibinfo{publisher}{IEEE}, \bibinfo{year}{2005}).

\bibitem{intel_strained_si_enh}
\bibinfo{author}{Ghani, T.} \emph{et~al.}
\newblock \bibinfo{title}{A 90nm high volume manufacturing logic technology featuring novel 45nm gate length strained silicon {CMOS} transistors}.
\newblock In \emph{\bibinfo{booktitle}{2003 IEDM}}, \bibinfo{pages}{978--980} (\bibinfo{year}{2003}).

\bibitem{taseen_tmd_strain}
\bibinfo{author}{Afrid, S. M. T.-S.}, \bibinfo{author}{Zhao, H.~L.}, \bibinfo{author}{van~der Zande, A.~M.} \& \bibinfo{author}{Rakheja, S.}
\newblock \bibinfo{title}{Strain-tunable inter-valley scattering defines universal mobility enhancement in n- and p-type {2D} {TMDs}} (\bibinfo{year}{2025}).
\newblock \urlprefix\url{https://www.arxiv.org/abs/2511.08975}.
\newblock \eprint{2511.08975}.

\bibitem{wiktor2016absolute}
\bibinfo{author}{Wiktor, J.} \& \bibinfo{author}{Pasquarello, A.}
\newblock \bibinfo{title}{Absolute deformation potentials of two-dimensional materials}.
\newblock \emph{\bibinfo{journal}{Phys. Rev. B}} \textbf{\bibinfo{volume}{94}}, \bibinfo{pages}{245411} (\bibinfo{year}{2016}).

\bibitem{cheng2020using}
\bibinfo{author}{Cheng, X.} \emph{et~al.}
\newblock \bibinfo{title}{Using strain to alter the energy bands of the monolayer {MoSe\textsubscript{2}}: A systematic study covering both tensile and compressive states}.
\newblock \emph{\bibinfo{journal}{Appl. Surf. Sci.}} \textbf{\bibinfo{volume}{521}}, \bibinfo{pages}{146398} (\bibinfo{year}{2020}).

\bibitem{shen2016strain}
\bibinfo{author}{Shen, T.}, \bibinfo{author}{Penumatcha, A.~V.} \& \bibinfo{author}{Appenzeller, J.}
\newblock \bibinfo{title}{Strain engineering for transition metal dichalcogenides based field effect transistors}.
\newblock \emph{\bibinfo{journal}{ACS Nano}} \textbf{\bibinfo{volume}{10}}, \bibinfo{pages}{4712--4718} (\bibinfo{year}{2016}).

\bibitem{leveillee2023ab}
\bibinfo{author}{Leveillee, J.}, \bibinfo{author}{Zhang, X.}, \bibinfo{author}{Kioupakis, E.} \& \bibinfo{author}{Giustino, F.}
\newblock \bibinfo{title}{\textit{Ab initio} calculation of carrier mobility in semiconductors including ionized-impurity scattering}.
\newblock \emph{\bibinfo{journal}{Phys. Rev. B}} \textbf{\bibinfo{volume}{107}}, \bibinfo{pages}{125207} (\bibinfo{year}{2023}).

\bibitem{zhang2022phonon}
\bibinfo{author}{Zhang, C.} \& \bibinfo{author}{Liu, Y.}
\newblock \bibinfo{title}{Phonon-limited transport of two-dimensional semiconductors: Quadrupole scattering and free carrier screening}.
\newblock \emph{\bibinfo{journal}{Phys. Rev. B}} \textbf{\bibinfo{volume}{106}}, \bibinfo{pages}{115423} (\bibinfo{year}{2022}).

\bibitem{pena_moire}
\bibinfo{author}{Peña, T.} \emph{et~al.}
\newblock \bibinfo{title}{{Moiré} engineering in {2D} heterostructures with process-induced strain}.
\newblock \emph{\bibinfo{journal}{Appl. Phys. Lett.}} \textbf{\bibinfo{volume}{122}} (\bibinfo{year}{2023}).

\bibitem{slidetronics}
\bibinfo{author}{Sui, F.} \emph{et~al.}
\newblock \bibinfo{title}{Atomic-level polarization reversal in sliding ferroelectric semiconductors}.
\newblock \emph{\bibinfo{journal}{Nature Comm.}} \textbf{\bibinfo{volume}{15}}, \bibinfo{pages}{1--8} (\bibinfo{year}{2024}).

\bibitem{stressor_engineering}
\bibinfo{author}{Liu, Q.} \emph{et~al.}
\newblock \bibinfo{title}{Programmable strainscapes in a two-dimensional ({2D}) material monolayer}.
\newblock \emph{\bibinfo{journal}{ACS Nano}} \textbf{\bibinfo{volume}{19}}, \bibinfo{pages}{30125--30136} (\bibinfo{year}{2025}).

\bibitem{chen_mos2_enh}
\bibinfo{author}{Chen, Y.} \emph{et~al.}
\newblock \bibinfo{title}{Mobility enhancement of strained {MoS\textsubscript{2}} transistor on flat substrate}.
\newblock \emph{\bibinfo{journal}{ACS Nano}} \textbf{\bibinfo{volume}{17}}, \bibinfo{pages}{14954--14962} (\bibinfo{year}{2023}).

\bibitem{gold_exfol_nature}
\bibinfo{author}{Huang, Y.} \emph{et~al.}
\newblock \bibinfo{title}{Universal mechanical exfoliation of large-area 2d crystals}.
\newblock \emph{\bibinfo{journal}{Nature Communications 2020 11:1}} \textbf{\bibinfo{volume}{11}}, \bibinfo{pages}{1--9} (\bibinfo{year}{2020}).

\bibitem{giannozzi2017advanced}
\bibinfo{author}{Giannozzi, P.} \emph{et~al.}
\newblock \bibinfo{title}{Advanced capabilities for materials modelling with {Quantum ESPRESSO}}.
\newblock \emph{\bibinfo{journal}{J. Phys.: Condens. Matter}} \textbf{\bibinfo{volume}{29}}, \bibinfo{pages}{465901} (\bibinfo{year}{2017}).

\bibitem{perdew1996generalized}
\bibinfo{author}{Perdew, J.~P.}, \bibinfo{author}{Burke, K.} \& \bibinfo{author}{Wang, Y.}
\newblock \bibinfo{title}{Generalized gradient approximation for the exchange-correlation hole of a many-electron system}.
\newblock \emph{\bibinfo{journal}{Phys. Rev. B}} \textbf{\bibinfo{volume}{54}}, \bibinfo{pages}{16533} (\bibinfo{year}{1996}).

\bibitem{giustino2017electron}
\bibinfo{author}{Giustino, F.}
\newblock \bibinfo{title}{Electron-phonon interactions from first principles}.
\newblock \emph{\bibinfo{journal}{Rev. Mod. Phys.}} \textbf{\bibinfo{volume}{89}}, \bibinfo{pages}{015003} (\bibinfo{year}{2017}).

\bibitem{baroni2001phonons}
\bibinfo{author}{Baroni, S.}, \bibinfo{author}{De~Gironcoli, S.}, \bibinfo{author}{Dal~Corso, A.} \& \bibinfo{author}{Giannozzi, P.}
\newblock \bibinfo{title}{Phonons and related crystal properties from density-functional perturbation theory}.
\newblock \emph{\bibinfo{journal}{Rev. Mod. Phys.}} \textbf{\bibinfo{volume}{73}}, \bibinfo{pages}{515} (\bibinfo{year}{2001}).

\bibitem{lundstrom2002fundamentals}
\bibinfo{author}{Lundstrom, M.}
\newblock \bibinfo{title}{Fundamentals of carrier transport, 2nd edn}.
\newblock \emph{\bibinfo{journal}{Meas. Sci. Technol.}} \textbf{\bibinfo{volume}{13}}, \bibinfo{pages}{230--230} (\bibinfo{year}{2002}).

\bibitem{stoney_eq}
\bibinfo{author}{Stoney, G.~G.}
\newblock \bibinfo{title}{The tension of metallic films deposited by electrolysis}.
\newblock \emph{\bibinfo{journal}{Proceedings of the Royal Society of London. Series A, Containing Papers of a Mathematical and Physical Character}} \textbf{\bibinfo{volume}{82}}, \bibinfo{pages}{172--175} (\bibinfo{year}{1909}).
\newblock \eprint{https://royalsocietypublishing.org/rspa/article-pdf/82/553/172/29858/rspa.1909.0021.pdf}.

\end{thebibliography}


\clearpage
\section*{Appendix: Extended Data Figures}

\FloatBarrier
\setcounter{figure}{0}
\renewcommand{\thefigure}{X\arabic{figure}} 

\begin{figure}[ht]
    \centering
    \includegraphics[width=0.5\linewidth]{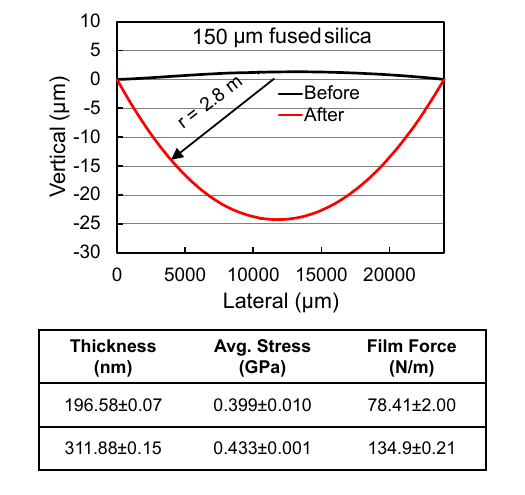}
    \caption{\textbf{Thin-film force measurement via bending.} Plots of substrate deflection of a 150 \textgreek{m}m fused silica substrate before and after depositing 100 nm of {AlO\textsubscript{x}}\ via e-beam evaporation at 0.5 \AA/s. The upward curvature is consistent with deposition of a film under high tensile stress. Applying the Stoney equation \cite{stoney_eq} yielded the film stress and film forces displayed in the bottom table.}
    \label{fig:ex_wafer_bend}
\end{figure}

\begin{figure}[ht]
    \centering
    \includegraphics[width=1\linewidth]{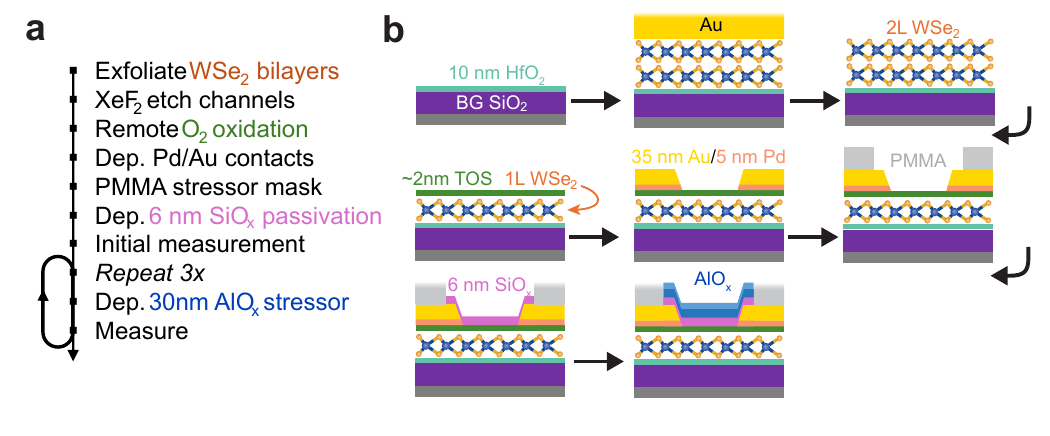}
    \caption{\textbf{Fabrication flow illustration.} \textbf{a,} Brief summary of key nanofabrication steps for the strained {WSe\textsubscript{2}}\ FETs. Details of each step is in the methods. \textbf{b,} Cross-section profile illustrations of changes in structure through process flow. The final step shows the end result after several iterations of stressor deposition and measurement.}
    \label{fig:ex_fab}
\end{figure}

\begin{figure}[ht]
    \centering
    \includegraphics[width=0.5\linewidth]{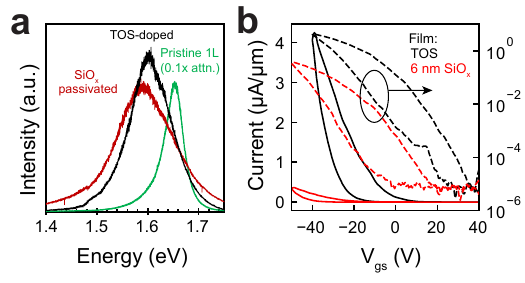}
    \caption{\textbf{Effects of TOS doping and {SiO\textsubscript{x}}\ passivation on WSe\textsubscript{2} p-FET.} \textbf{a,} Photoluminescence spectra comparing a pristine monolayer {WSe\textsubscript{2}}\ from exfoliation to monolayer {WSe\textsubscript{2}}\/TOS heterostructure formed by remote oxidization of bilayer {WSe\textsubscript{2}} and a channel region after SiO\textsubscript{x} passivation. The monolayer WSe\textsubscript{2}/TOS heterostructure had a 10\texttimes\ attenuation in peak intensity and increased the FWHM compared with the exfoliated monolayer. These changes are consistent with increasing hole concentration.\cite{sihan_TOS} Crucially, the {SiO\textsubscript{x}}\ passivation did not significantly attenuate the peak intensity or increase the FWHM, indicating no significant change in strain, hole concentration, or defect density through fabrication of the doped monolayer into a p-FET. \textbf{b,} Current density versus gate voltage transfer characteristics comparing the TOS p-doped {WSe\textsubscript{2}}\ FETs before and after the {SiO\textsubscript{x}}\ passivation. The overall current density reduced by 10x, while the V\textsubscript{T} and hysteresis did not significantly change.}
    \label{fig:ex_doping_pasiv}
\end{figure}

 \begin{figure}[ht]
    \centering
    \includegraphics[width=1\linewidth]{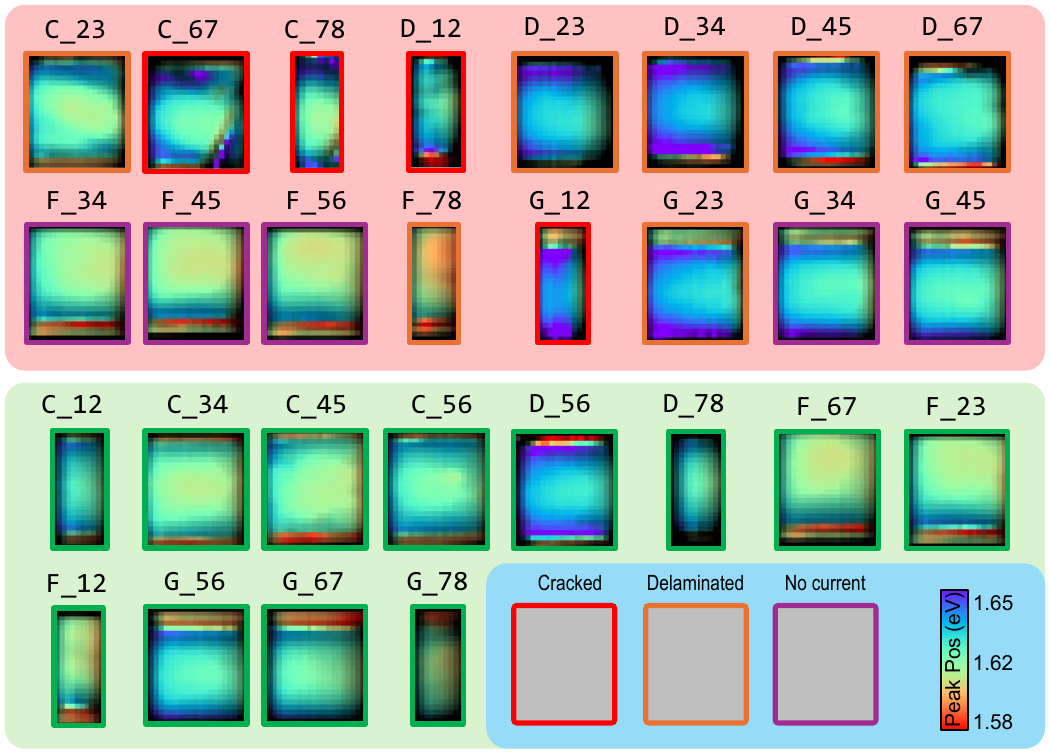}
    \caption{\textbf{Hyperspectral maps of photoluminescence peak energy for all measured p-FETs with 90 nm {AlO\textsubscript{x}}\ stressor.} The hue corresponds to PL peak energy, while the brightness corresponds to PL peak intensity. The letters A-G indicate different FET arrays, while the numbers indicate different FETs on each array. FET arrays A and B suffered gate breakdown during initial measurements and were thus left out of subsequent measurements.    
    }
    \label{fig:ex_reject}
\end{figure}

\begin{figure}[ht]
    \centering
    \includegraphics[width=1\linewidth]{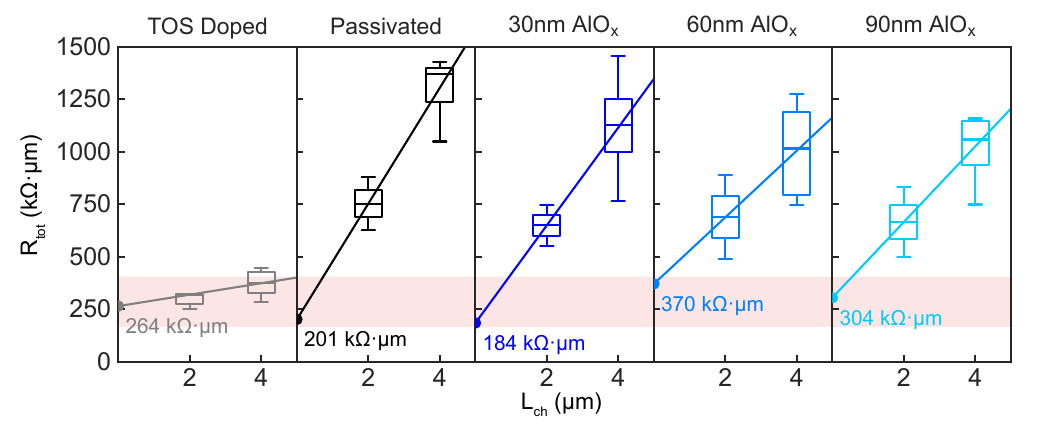}
    \caption{\textbf{Transfer length measurements to estimate contact resistance.} Plots of total resistivity (R\textsubscript{tot}) of FETs at V\textsubscript{OD}~=~-20~V with L\textsubscript{ch}~=~2~\textgreek{m}m and 4~\textgreek{m}m, linearly extrapolated to obtain 2R\textsubscript{C} estimations. The values of 2R\textsubscript{C} range from 200-370 k\textgreek{W}. These values are significant compared with some of the best contact resistances demonstrated. However, the contacts remain ohmic and small compared with the sheet resistance. The relative value of 2R\textsubscript{C} does not show a significant trend as a function of stressor thickness, indicating that the extracted performance parameters in this study are channel-dominated.}
    \label{fig:ex_TLM}
\end{figure}

\begin{figure}[ht]
    \centering
    \includegraphics[width=1\linewidth]{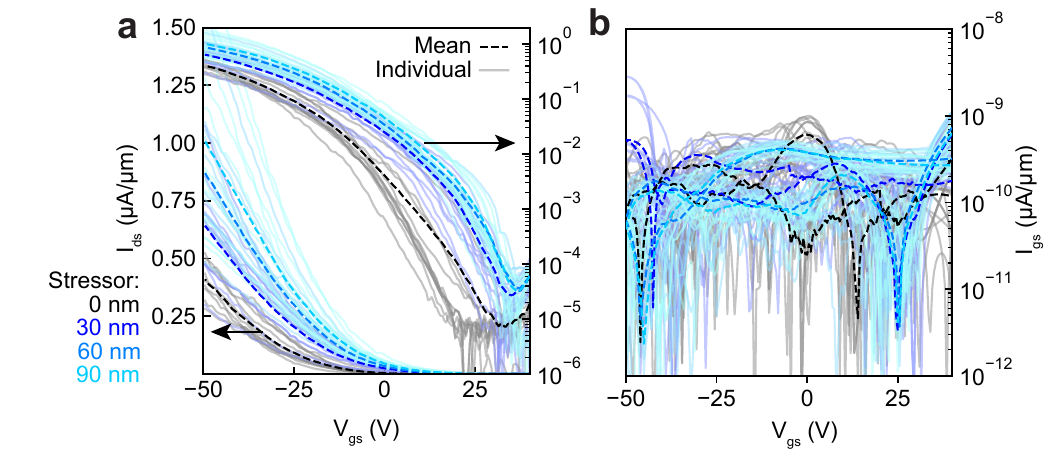}
    \caption{\textbf{Comparison of transfer curves and gate leakage.} \textbf{a,} Plot of I\textsubscript{ds}--V\textsubscript{gs} transfer characteristics of all 12 FETs at V\textsubscript{ds}~=~0.5~V, for each stressor thickness. ``0 nm" refers to FETs after initial {SiO\textsubscript{x}}\ passivation. The dotted lines show the mean curves at each stressor thickness. \textbf{b,} Plot of corresponding I\textsubscript{gs}--V\textsubscript{gs} leakage currents throughout stressor iterations.}
    \label{fig:ex_leakage}
\end{figure}

\begin{figure}[ht]
    \centering
    \includegraphics[width=0.97\linewidth]{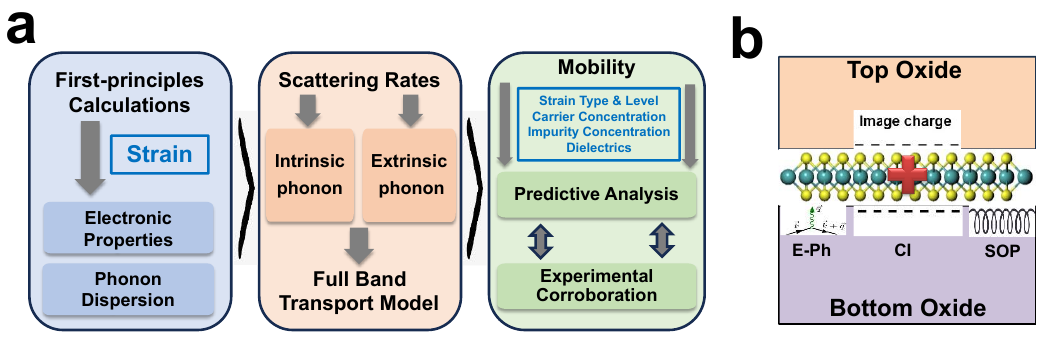}
    \caption{\textbf{Multiscale simulation framework and scattering processes governing charge transport in 2D TMDs.} 
    \textbf{a,} Workflow of the multiscale computational approach used to predict carrier mobility, beginning with first-principles calculations of electronic structure and phonon properties under varying strain. These outputs are used to compute scattering rates from intrinsic phonons (ADP, ODP, POP, IV, PZ) and extrinsic phonons (CI, SO), which are incorporated into a full-band transport model to obtain the strain-, carrier-, impurity-, and dielectric-dependent mobility. 
    \textbf{b,} Schematic overview of the dominant scattering mechanisms in monolayer TMDs, including electron-phonon interactions driven by lattice vibrations, Coulomb scattering from charged impurities, and remote phonon coupling with polar optical modes of surrounding oxide layers.}
    \label{fig:sim_flow}
\end{figure}

\end{document}


\selectlanguage{english}
\title[Article Title]{Supplementary Information: Enhancing Hole Mobility in Monolayer \WSe\ p-FETs via Process-Induced Compression}
\author[1,4,5,6]{\fnm{He Lin} \sur{Zhao}}

\author[1,5,6]{\fnm{Sheikh Mohd Ta-Seen} \sur{Afrid}}

\author[3,4,6]{\fnm{Dongyoung} \sur{Yoon}}

\author[2,4,6]{\fnm{Zachary} \sur{Martin}}

\author[3,4,6]{\fnm{Zakaria} \sur{Islam}}

\author[3,4,6]{\fnm{Sihan} \sur{Chen}}

\author[3,4,6]{\fnm{Yue} \sur{Zhang}}

\author[2,4,6]{\fnm{Pinshane Y.} \sur{Huang}}

\author[1,5,6]{\fnm{Shaloo} \sur{Rakheja}}

\author*[1,2,3,4,5,6]{\fnm{Arend M.} \sur{van der Zande}}\email{arendv@illinois.edu}

\affil[1]{\orgdiv{Department of Electrical and Computer Engineering}, \orgname{University of Illinois Urbana-Champaign}, \orgaddress{\city{Urbana}, \postcode{61801}, \state{Illinois}, \country{United States}}}

\affil[2]{\orgdiv{Department of Materials Science and Engineering}, \orgname{University of Illinois Urbana-Champaign}, \orgaddress{\city{Urbana}, \postcode{61801}, \state{Illinois}, \country{United States}}}

\affil[3]{\orgdiv{Department of Mechanical Science and Engineering}, \orgname{University of Illinois Urbana-Champaign}, \orgaddress{\city{Urbana}, \postcode{61801}, \state{Illinois}, \country{United States}}}

\affil[4]{\orgdiv{Materials Research Laboratory}, \orgname{University of Illinois Urbana-Champaign}, \orgaddress{\city{Urbana}, \postcode{61801}, \state{Illinois}, \country{United States}}}

\affil[5]{\orgdiv{Holonyak Micro and Nanotechnology Laboratory}, \orgname{University of Illinois Urbana-Champaign}, \orgaddress{\city{Urbana}, \postcode{61801}, \state{Illinois}, \country{United States}}}

\affil[6]{\orgdiv{Grainger College of Engineering}, \orgname{University of Illinois Urbana-Champaign}, \orgaddress{\city{Urbana}, \postcode{61801}, \state{Illinois}, \country{United States}}}

\maketitle

\renewcommand{\thetable}{S\arabic{table}}
\renewcommand{\thefigure}{S\arabic{figure}}

\FloatBarrier
\section{Passivation Strategies}
\FloatBarrier
\normalsize
Properly passivating and encapsulating the surface of 2D FETs is critical for minimizing damage to the channel and maintaining doping concentration. \cite{hfo2_passivation, mos2_alox_enh, al2o3_ndope}
Previously, we found that the direct deposition of the stressor damaged MoS\textsubscript{2} n-FETs, leading to poor performance, and that an ALD passivation layer was important for maintaining performance.\cite{yue_mos2_enh} 
Thus, a key enabling step in this study was to determine a strategy to passivate the p-FETs while maintaining the integrity of WSe\textsubscript{2} and the p-doping provided by the TOS layer. 
We compared different passivation layer strategies to enable this study, including no passivation layer, Atomic Layer deposition (ALD) of Al\textsubscript{2}O\textsubscript{3} or HfO\textsubscript{2}, and the low-power e-beam evaporation of \SiOx. 
In each case, we kept the thickness of the passivation layer to a few nm, so it would be enough to protect the underlying 2D material, but not thick enough to suppress the strain transfer during stressor deposition.\par
We found that the direct deposition of \AlOx\ rendered most FETs non-functional. We hypothesize that the high energy required for \mbox{Al\textsubscript{2}O\textsubscript{3}} evaporation damaged the 2D channel, as supported by drastically attenuated PL intensities. 
In addition, ALD HfO\textsubscript{2} and Al\textsubscript{2}O\textsubscript{3} passivation caused n-doping in the FETs so severe that they could not reach the threshold voltage, rendering meaningful transport characterization impossible.
We hypothesize that this is a combination of the n-doping effects of ALD \mbox{Al\textsubscript{2}O\textsubscript{3}} or \mbox{HfO\textsubscript{2}} \cite{hfo2_passivation, al2o3_ndope} and the presence of \mbox{H\textsubscript{2}O} precursor, which is known to degrade the TOS layer. \cite{borah_TOS_doping}
Of all the methods we tried, e-beam evaporated \SiOx\ required the least energy and also contained no water; in addition, it caused no discernible decrease in PL intensity, indicating little to no direct effects on the integrity of the WSe\textsubscript{2} at the molecular level.
The e-beam evaporated \SiOx\ passivation process did degrade mobility on average by a factor of 3.5. However, of the techniques we tried, it was the only one that maintained the doping level and PL intensity, as shown in Extended Data Fig. X2. \par
Due to the stochastic and non-conformal nature of e-beam evaporation, special care was needed to ensure that the deposited layer did not form granular crystals. Should granular crystals be present, we found that the subsequent stressor layers would preferentially deposit at the sites of the grains, creating pillars instead of uniform films. Such stressor layers damaged the WSe\textsubscript{2} and produced no strain. To prevent granular deposition, we maximized evaporant diffusion time by lowering the e-beam power to the minimum possible, yielding a deposition rate of 0.03 \AA/s. The surface topographies of the finalized recipes are shown in Supplementary Fig. \ref{fig:sup_AFM}. We note that during the lithography steps, some PMMA residue was present. We limited cleaning steps due to the water and temperature sensitivity of the TOS doping layer. There was no evidence that the residue negatively impacted the uniformity of the subsequent \SiOx\ passivation and \AlOx\ stressor films.

\begin{figure}[ht]
    \centering
    \includegraphics[width=0.5\linewidth]{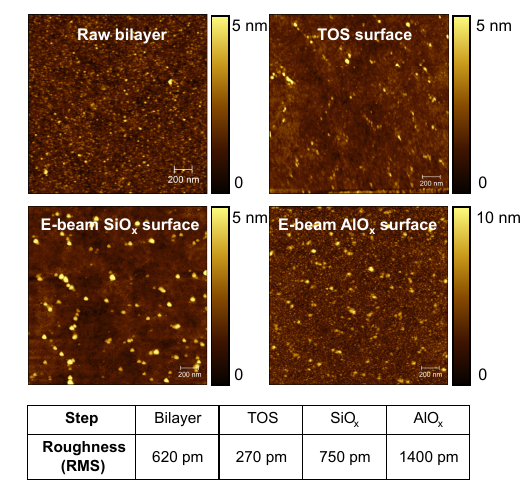}
    \caption{\textbf{Surface roughness of \WSe\ channels through fabrication process.} AFM scans and roughness of a representative \WSe\ channel region through relevant nanofabrication steps.}
    \label{fig:sup_AFM}
\end{figure}

\FloatBarrier
\section{Finite-Element Analysis}
\FloatBarrier
Below are the detailed parameters for the ABAQUS FEA simulations.

\begin{table}[h]
    \small
    \begin{tabular}{|c|cccc|}
    \hline
        & \makecell{\AlOx \\ Stressor} & \makecell{Au \\ Contacts} & \makecell{SiO\textsubscript{2} \\ Substrate} & \makecell{\WSe \\ Channel} \\ \hline
        \makecell{Lateral \\ dimensions \\ (\textgreek{m}m)} & 4\texttimes5 & 5\texttimes5 & 12\texttimes12 & 4.4\texttimes5 \\ \hline
        \makecell{Thickness \\ (nm)} & 50 & 50 & 100 & 10 \\ \hline
        \makecell{Young's \\ modulus \\ (GPa)} & 250 & 79 & 57 & 265 \\ \hline
        \makecell{Poisson's \\ ratio} & 0.18 & 0.42 & 0.17 & 0.25 \\ \hline
    \end{tabular}
    \caption{\textbf{Parameters used in FEA simulation.} The simulation setup is illustrated in Main Figure 2b. Lateral dimensions are reported in the format of $x \times y$, where $x$ is the longitudinal channel dimension and $y$ is the lateral channel dimension.}
    \label{tab:FEA_params}
\end{table}

\FloatBarrier
\section{Photoluminescence-based FET Device Health Decision}
\FloatBarrier
Based on the PL maps shown in Extended Data Fig. X5, we accepted FETs that yielded PL peak maps that showed no obvious discontinuities and produced current under voltage bias. FETs were rejected for primarily three reasons: channel damage, delamination, or failure to conduct. 
Channel damage appeared as inhomogeneities or discontinuities in PL peak energies. Channel delamination manifested as either a sharp PL blueshift or a significant intensity dimming at the edges. 
Lastly, some FET channels appeared acceptable under PL mapping, but ceased to conduct after a certain stressor thickness was reached. We hypothesize that these FETs had nanoscale cracks that were too small to be visible through optical imaging, eventually leading to total channel failure.

\FloatBarrier
\section{Transport Parameter Extraction}
\FloatBarrier

 We use linear transfer curve fitting to extract transport parameters in this project. We extract the threshold voltage $V_\text{T}$ by performing linear regression on the drain-source current $I_\text{ds}$ as a function of gate voltage $V_\text{gs}$ about the neighborhood of the point of peak transconductance $g_m$, which we refer to as $V_{\text{lin}}$. The intercept of the fitted line with the 0-current axis is then $V_\text{T}$:
\begin{equation}
    V_\text{T} = V_{\text{lin}} - \frac{I_{\text{lin}}}{\frac{\partial I_{\text{ds}}}{\partial V_{\text{gs}}}\big\rvert_{V_{\text{gs}}=V_{\text{lin}}}}
\end{equation}
And the field-effect mobility $\mu_\text{FE}$ is a function of the slope of the fitted line:
\begin{equation}
    \mu_\text{FE} = \left(\frac{L}{W
V_{\text{ds}} C_{\text{ox}}}\right)
\frac{\partial I_{\text{ds}}}{\partial V_{\text{gs}}} \bigg\rvert_{V_{\text{gs}}=V_{\text{lin}}}
\end{equation}
    
Where $L,W$ are the length (4 or 2~\textgreek{m}m) and width (5~\textgreek{m}m) of the \WSe\ channel, $V_\text{ds}$ is the drain-source voltage. We calculate the gate dielectric capacitance $C_\text{ox}$ to be \mbox{38.4 nF/cm\textsuperscript{2}} by using an oxide thickness of 90~nm and the dielectric constant of SiO\textsubscript{2}, 3.9. The series capacitance of the 10~nm HfO\textsubscript{2} etch-stop is significantly larger, so its effect on the overall capacitance is negligible.
We extract the maximum current output density $I_{\text{on}}$ at a constant overdrive voltage $V_\text{OD}$ (--20~V) to exclude doping and hysteresis effects:
\begin{equation}
    I_\text{on} = I_\text{ds}(V_\text{gs})\bigg\rvert_{V_\text{gs} = V_\text{T} - 20 \text{V}}
\end{equation}

Supplementary Tables \ref{tab:mu_relative}-\ref{tab:VT_raw} provide the extracted absolute and relative values of $\mu_\text{FE}$, $I_\text{on}$, and $V_\text{T}$ for all 12 acceptable FETs at each measured thickness. 

\begin{table}[h]
    \small 
    \begin{tabular}{|c|cccc|}
    \hline
        Device & 0 nm & 30 nm & 60 nm & 90 nm \\ \hline
        C\_12 & 1 & 1.31 & 1.41 & 1.45 \\ \hline
        C\_34 & 1 & 1.57 & 1.71 & 1.41 \\ \hline
        C\_45 & 1 & 1.21 & 1.75 & 1.42 \\ \hline
        C\_56 & 1 & 1.33 & 1.32 & 1.41 \\ \hline
        D\_56 & 1 & 1.23 & 1.11 & 1.31 \\ \hline
        D\_78 & 1 & 1.78 & 2.72 & 3.22 \\ \hline
        F\_12 & 1 & N/A & 1.58 & 1.99 \\ \hline
        F\_23 & 1 & 1.23 & 1.11 & 1.51 \\ \hline
        F\_67 & 1 & 1.10 & 1.19 & 1.42 \\ \hline
        G\_56 & 1 & 1.41 & 1.57 & 2.17 \\ \hline
        G\_67 & 1 & 1.35 & 1.79 & 1.88 \\ \hline
        G\_78 & 1 & 1.18 & 1.38 & 1.42 \\ \hline
    \end{tabular}
    \caption{\textbf{Table of relative mobility factors.} The thickness refers to AlO\textsubscript{x} stressor thickness. The N/A represents a bad measurement that did not yield usable data.}
    \label{tab:mu_relative}
\end{table}

\begin{table}[h]
    \small 
    \begin{tabular}{|c|cccc|}
    \hline
        Device & 0 nm & 30 nm & 60 nm & 90 nm \\ \hline
        C\_12 & 1.46 & 1.91 & 2.06 & 2.12 \\ \hline
        C\_34 & 2.36 & 3.72 & 4.05 & 3.32 \\ \hline
        C\_45 & 2.23 & 2.69 & 3.89 & 3.16 \\ \hline
        C\_56 & 2.04 & 2.72 & 2.7 & 2.88 \\ \hline
        D\_56 & 3.73 & 4.6 & 4.15 & 4.9 \\ \hline
        D\_78 & 1.46 & 2.6 & 3.97 & 4.7 \\ \hline
        F\_12 & 1.85 & N/A & 2.93 & 3.69 \\ \hline
        F\_23 & 3.51 & 4.31 & 3.91 & 5.29 \\ \hline
        F\_67 & 3.46 & 3.83 & 4.11 & 4.93 \\ \hline
        G\_56 & 3.94 & 5.56 & 6.2 & 8.56 \\ \hline
        G\_67 & 3.56 & 4.79 & 6.37 & 6.68 \\ \hline
        G\_78 & 3.76 & 4.42 & 5.2 & 5.34 \\ \hline
    \end{tabular}
    \caption{\textbf{Table of field-effect mobilities in cm\textsuperscript{2}/Vs.} The thickness refers to AlO\textsubscript{x} stressor thickness. The N/A represents a bad measurement that did not yield usable data.}
    \label{tab:mu_raw}
\end{table}

\begin{table}[h]
    \small 
    \begin{tabular}{|c|cccc|}
    \hline
        Device & 0 nm & 30 nm & 60 nm & 90 nm \\ \hline
        C\_12 & 1 & 1.09 & 1.24 & 1.28 \\ \hline
        C\_34 & 1 & 1.50 & 1.71 & 1.39 \\ \hline
        C\_45 & 1 & 1.36 & 1.76 & 1.54 \\ \hline
        C\_56 & 1 & 1.21 & 1.11 & 1.20 \\ \hline
        D\_56 & 1 & 1.22 & 1.18 & 1.50 \\ \hline
        D\_78 & 1 & 1.42 & 2.29 & 2.71 \\ \hline
        F\_12 & 1 & N/A & 1.59 & 2.05 \\ \hline
        F\_23 & 1 & 1.12 & 1.11 & 1.47 \\ \hline
        F\_67 & 1 & 1.11 & 1.14 & 1.35 \\ \hline
        G\_56 & 1 & 1.55 & 1.70 & 2.37 \\ \hline
        G\_67 & 1 & 1.35 & 1.81 & 1.88 \\ \hline
        G\_78 & 1 & 2.05 & 2.32 & 2.26 \\ \hline
    \end{tabular}
    \caption{\textbf{Table of on-current factors.} The thickness refers to AlO\textsubscript{x} stressor thickness. The N/A represents a bad measurement that did not yield usable data.}
    \label{tab:Ion_relative}
\end{table}

\begin{table}[h]
    \small 
    \begin{tabular}{|c|cccc|}
    \hline
        Device & 0 nm & 30 nm & 60 nm & 90 nm \\ \hline
        C\_12 & 0.326 & 0.356 & 0.405 & 0.419 \\ \hline
        C\_34 & 0.245 & 0.368 & 0.419 & 0.340 \\ \hline
        C\_45 & 0.207 & 0.280 & 0.364 & 0.319 \\ \hline
        C\_56 & 0.227 & 0.274 & 0.252 & 0.272 \\ \hline
        D\_56 & 0.361 & 0.439 & 0.427 & 0.540 \\ \hline
        D\_78 & 0.408 & 0.509 & 0.425 & 0.442 \\ \hline
        F\_12 & 0.393 & N/A & 0.626 & 0.807 \\ \hline
        F\_23 & 0.354 & 0.395 & 0.393 & 0.519 \\ \hline
        F\_67 & 0.350 & 0.388 & 0.400 & 0.471 \\ \hline
        G\_56 & 0.362 & 0.561 & 0.617 & 0.857 \\ \hline
        G\_67 & 0.356 & 0.480 & 0.645 & 0.669 \\ \hline
        G\_78 & 0.443 & 0.908 & 1.028 & 1.002 \\ \hline
    \end{tabular}
    \caption{\textbf{Table of absolute on-current in \mbox{\textgreek{m}A/\textgreek{m}m}.} The thickness refers to AlO\textsubscript{x} stressor thickness. The N/A represents a bad measurement that did not yield usable data.}
    \label{tab:Ion_raw}
\end{table}

\begin{table}[h]
    \small 
    \begin{tabular}{|c|cccc|}
    \hline
        Device & 0 nm & 30 nm & 60 nm & 90 nm \\ \hline
        C\_12 & 0 & 15.43 & 10.09 & 12.02 \\ \hline
        C\_34 & 0 & 7.68 & 9.80 & 14.56 \\ \hline
        C\_45 & 0 & 26.44 & 7.58 & 6.86 \\ \hline
        C\_56 & 0 & 9.22 & 10.95 & 14.21 \\ \hline
        D\_56 & 0 & -3.19 & 13.92 & 8.30 \\ \hline
        D\_78 & 0 & 6.70 & 3.88 & 11.02 \\ \hline
        F\_12 & 0 & N/A & -0.28 & 7.22 \\ \hline
        F\_23 & 0 & 5.54 & 9.79 & 9.84 \\ \hline
        F\_67 & 0 & 15.63 & 12.56 & 11.30 \\ \hline
        G\_56 & 0 & 9.24 & 13.36 & 17.65 \\ \hline
        G\_67 & 0 & 13.39 & 14.64 & 14.81 \\ \hline
        G\_78 & 0 & -10.84 & -15.05 & -12.82 \\ \hline
    \end{tabular}
    \caption{\textbf{Table of threshold voltage shifts.} The thickness refers to AlO\textsubscript{x} stressor thickness. The N/A represents a bad measurement that did not yield usable data.}
    \label{tab:VT_relative}
\end{table}

\begin{table}[h]
    \small 
    \begin{tabular}{|c|cccc|}
    \hline
        Device & 0 nm & 30 nm & 60 nm & 90 nm \\ \hline
        C\_12 & -16.77 & -1.34 & -6.68 & -4.75 \\ \hline
        C\_34 & -19.18 & -11.5 & -9.38 & -4.62 \\ \hline
        C\_45 & -14.78 & 11.67 & -7.19 & -7.92 \\ \hline
        C\_56 & -18.44 & -9.22 & -7.49 & -4.24 \\ \hline
        D\_56 & -13.33 & -16.52 & 0.6 & -5.02 \\ \hline
        D\_78 & -26.68 & -19.98 & -22.79 & -15.65 \\ \hline
        F\_12 & -2.61 & N/A & -2.89 & 4.61 \\ \hline
        F\_23 & -21.88 & -16.34 & -12.09 & -12.04 \\ \hline
        F\_67 & -23.09 & -7.45 & -10.53 & -11.78 \\ \hline
        G\_56 & -24.75 & -15.51 & -11.39 & -7.11 \\ \hline
        G\_67 & -25.71 & -12.32 & -11.07 & -10.89 \\ \hline
        G\_78 & 7.04 & -10.84 & -15.05 & -12.82 \\ \hline
    \end{tabular}
    \caption{\textbf{Table of absolute threshold voltages.} The thickness refers to AlO\textsubscript{x} stressor thickness. The N/A represents a bad measurement that did not yield usable data.}
    \label{tab:VT_raw}
\end{table}

\FloatBarrier
\section{DFT and Transport Simulation Data}
\FloatBarrier
The following are the parameters extracted from first-principles calculations and transport simulations.

\begin{table}[h]
    \small
    \begin{tabular}{|c|cc|}
    \hline
    \makecell{Biaxial Strain \\ (\%$\varepsilon$)} & $E_{\Gamma}$ (meV) & $E_{K}$ (meV) \\ \hline

        -1.00  & -325 & -7.24 \\ \hline
        -0.75  & -295 & -6.10 \\ \hline
        -0.50  & -260 & -5.95 \\ \hline
        -0.25  & -220 & -5.25 \\ \hline
         0.00  & -175 & -4.84 \\ \hline
         0.25  & -147 & -4.32 \\ \hline
         0.50  & -118 & -3.89 \\ \hline
         0.75  & -95  & -2.76 \\ \hline
         1.00  & -70  & -2.54 \\ \hline
    \end{tabular}
    \caption{\textbf{Table of valence band maxima (VBM) energies at $\Gamma$ and K points under biaxial strain.} 
    Energies extracted from Main Fig. 4b for biaxial strains from -1\% to +1\%.}
    \label{tab:gamma_k_energy}
\end{table}

\begin{table}[h]
    \small
    \begin{tabular}{|c|cc|}
    \hline
 \makecell{Biaxial Strain \\ (\%$\varepsilon$)} & \makecell{With IV scattering \\ ($\times 10^{12}$ s$^{-1}$)} & \makecell{Without IV scattering \\ ($\times 10^{12}$ s$^{-1}$)} \\ \hline

        -1.00  & 3.038 & 2.805 \\ \hline
        -0.75  & 3.338 & 2.907 \\ \hline
        -0.50  & 3.659 & 3.001 \\ \hline
        -0.25  & 3.964 & 3.097 \\ \hline
         0.00  & 4.200 & 3.188 \\ \hline
         0.25  & 4.629 & 3.274 \\ \hline
         0.50  & 5.075 & 3.366 \\ \hline
         0.75  & 5.617 & 3.468 \\ \hline
         1.00  & 6.137 & 3.570 \\ \hline
    \end{tabular}
    \caption{\textbf{Table of values with and without Inter-valley (IV) scattering rate under biaxial strain at 200 meV.} 
    Scattering rates extracted from Main Fig. 4c (right) for biaxial strains from -1\% to +1\%.}
    \label{tab:iv_scattering}
\end{table}

\begin{table}[h]
    \small
    \begin{tabular}{|c|cc|cc|}
    \hline
         \makecell{Biaxial Strain \\ (\%$\varepsilon$)} &
         \makecell{$\mu$ (Int.) \\ (cm$^2$/V$\cdot$s)} &
         \makecell{$\mu$ (Int. + Ext.) \\ (cm$^2$/V$\cdot$s)} &
         \makecell{$\mu/\mu_0$ \\ (Int.)} &
         \makecell{$\mu/\mu_0$ \\ (Int. + Ext.)} \\
    \hline
        -1.00  & 726.28 & 66.08 & 2.71 & 2.36 \\ \hline
        -0.75  & 659.28 & 60.76 & 2.46 & 2.17 \\ \hline
        -0.50  & 568.16 & 53.76 & 2.12 & 1.92 \\ \hline
        -0.25  & 450.24 & 42.28 & 1.68 & 1.51 \\ \hline
         0.00  & 268.00 & 28.00 & 1.00 & 1.00 \\ \hline
         0.25  & 214.40 & 18.20 & 0.80 & 0.65 \\ \hline
         0.50  & 168.84 & 13.16 & 0.63 & 0.47 \\ \hline
         0.75  & 142.04 & 9.52  & 0.53 & 0.34 \\ \hline
         1.00  & 120.60 & 7.28  & 0.45 & 0.26 \\ \hline
    \end{tabular}
    \caption{\textbf{Table of mobility and its enhancement factors for intrinsic (int.) and with extrinsic (ext.) effect under biaxial strain.}
    Mobility and its relative enhancements factor extracted from Main Fig. 4d. $\mu_0$ is the unstrained mobility.}
    \label{tab:enhancement_strain}
\end{table}

\begin{table}[h]
    \small
    \begin{tabular}{|c|ccc|}
    \hline
        & & $\mu/\mu_0$  & \\ \hline 
        \makecell{Biaxial Strain \\ (\%$\varepsilon$)} 
        & \makecell{$n_{\mathrm{imp}} =$ \\ $1\times10^{11}$ cm$^{-2}$}
        & \makecell{$n_{\mathrm{imp}} =$ \\ $5\times10^{12}$ cm$^{-2}$}
        & \makecell{$n_{\mathrm{imp}} =$ \\ $1\times10^{13}$ cm$^{-2}$} \\ \hline

        -1.00 & 3.14 & 2.96 & 2.74 \\ \hline
        -0.75 & 2.95 & 2.73 & 2.60 \\ \hline
        -0.50 & 2.65 & 2.41 & 2.28 \\ \hline
        -0.25 & 1.85 & 1.74 & 1.64 \\ \hline
        -0.20 & 1.67 & 1.59 & 1.47 \\ \hline
        -0.15 & 1.46 & 1.40 & 1.34 \\ \hline
        -0.10 & 1.34 & 1.28 & 1.20 \\ \hline
        -0.05 & 1.24 & 1.18 & 1.13 \\ \hline
         0.00 & 1.00 & 1.00 & 1.00 \\ \hline

    \end{tabular}
    \caption{\textbf{Table of mobility enhancement factors vs.~biaxial compressive strain 
    for different impurity concentrations ($n_{\mathrm{imp}}$)}. Enhancements extracted from Main Fig. 5a.}
    \label{tab:mobility_enhanc_impurity_strain}
\end{table}

\begin{table}[h]
    \small
    \centering
    \begin{tabular}{|c|c|c|c|}
    \hline
        \makecell{Carrier Conc. \\ (cm\textsuperscript{-2})}
        & \makecell{Impurity Conc. \\ (cm\textsuperscript{-2})}
        & Dielectric 
        &  \makecell{$\mu_{0}$ \\ (cm$^{2}$/V$\cdot$s)} \\ \hline
        
        $10^{11}$ & $2.5\times 10^{12}$ & SiO$_2$ & 12 \\ \hline
        $10^{12}$ & $2.5\times 10^{12}$ & SiO$_2$ & 17 \\ \hline
        $10^{13}$ & $2.5\times 10^{12}$ & SiO$_2$ & 28 \\ \hline
        
        $10^{13}$  & $10^{11}$ & SiO$_2$ & 183 \\ \hline
        $10^{13}$  & $10^{12}$ & SiO$_2$ & 28  \\ \hline
        $10^{13}$  & $10^{13}$ & SiO$_2$ & 13  \\ \hline
        
        $10^{13}$  & $2.5\times 10^{12}$ & SiO$_2$     & 28 \\ \hline
        $10^{13}$  & $2.5\times 10^{12}$ & Al$_2$O$_3$ & 32 \\ \hline
        $10^{13}$  & $2.5\times 10^{12}$ & HfO$_2$     & 18 \\ \hline
        
    \end{tabular}
    \caption{\textbf{Table of unstrained mobility under systematic variation of carrier concentration, impurity concentration, and dielectric environment.}  
    Unless otherwise specified, all calculations assume room temperature,  
    $p = 10^{13}$\,cm$^{-2}$, 
    $n_{\text{imp}} = 2.5\times 10^{12}$\,cm$^{-2}$, 
    and a SiO$_2$ dielectric and the data extracted from Main Fig. 5b.}
    \label{tab:mobility_wise}
\end{table}

\begin{table}[h]
    \small
    \centering
    \begin{tabular}{|c|c|c|ccc|}
    \hline
        \makecell{Carrier Conc. \\ (cm\textsuperscript{-2})}
        & \makecell{Impurity Conc. \\ (cm\textsuperscript{-2})} 
        & Dielectric
        & \makecell{$\frac{\partial(\mu/\mu_0)}{\partial \varepsilon}$ \\ (biaxial)}
        & \makecell{$\frac{\partial(\mu/\mu_0)}{\partial \varepsilon}$ \\ (armchair)}
        & \makecell{$\frac{\partial(\mu/\mu_0)}{\partial \varepsilon}$ \\ (zigzag)} \\ \hline

        $10^{11}$ & $2.5\times 10^{12}$ & SiO$_2$
        & 3.51 & 2.09 & 2.02 \\ \hline

        $10^{12}$ & $2.5\times 10^{12}$ & SiO$_2$
        & 3.88 & 2.29 & 2.18 \\ \hline

        $10^{13}$ & $2.5\times 10^{12}$ & SiO$_2$
        & 4.12 & 2.51 & 2.42 \\ \hline

        $10^{13}$ & $10^{11}$ & SiO$_2$
        & 4.39 & 2.88 & 2.84 \\ \hline

        $10^{13}$  & $10^{12}$ & SiO$_2$
        & 3.98 & 2.51 & 2.43 \\ \hline

        $10^{13}$  & $10^{13}$ & SiO$_2$
        & 3.68 & 2.10 & 1.95 \\ \hline

    $10^{13}$  & $2.5\times10^{12}$ & SiO$_2$
        & 4.18 & 2.54 & 2.41 \\ \hline

        $10^{13}$  & $2.5\times10^{12}$ & Al$_2$O$_3$
        & 4.20 & 2.61 & 2.48 \\ \hline

        $10^{13}$  & $2.5\times10^{12}$ & HfO$_2$
        & 3.92 & 2.45 & 2.34 \\ \hline

    \end{tabular}
    \caption{\textbf{Table of mobility enhancement factor per percent strain under biaxial, armchair, and zig-zag strain.}
    Default conditions (unless varied): 
    $p = 10^{13}$\,cm$^{-2}$, 
    $n_{\text{imp}} = 2.5\times10^{12}$\,cm$^{-2}$, 
    dielectric = SiO$_2$ and the data extracted from Main Fig. 5c.}
    \label{tab:strain_enhancement_wise}
\end{table}

\clearpage
